\documentclass[twocolumn, onecolumn]{article}
\usepackage[margin=1in,footskip=0.25in]{geometry}
\usepackage{authblk}
\usepackage{amsmath, bm}
\usepackage{amssymb}
\usepackage{graphicx}
\usepackage{hyperref}
\usepackage[linesnumbered,ruled,vlined]{algorithm2e}
\usepackage{float}

\usepackage{caption}
\usepackage{subcaption}
\usepackage{dblfloatfix}% http://ctan.org/pkg/dblfloatfix
\usepackage{lipsum}% http://ctan.org/pkg/lipsum

\DeclareMathOperator*{\argmin}{argmin}

\title{Dock2D: Synthetic data for the molecular recognition problem}
\author[1]{Siddharth Bhadra-Lobo\footnote{Equal contributors}\footnote{sid.bl@rutgers.edu}}
\author[2]{Georgy Derevyanko$^*$\footnote{georgy.derevyanko@gmail.com}}
\author[1, 3]{Guillaume Lamoureux\footnote{guillaume.lamoureux@rutgers.edu}}

\affil[1]{Center for Computational and Integrative Biology, Rutgers University, Camden, NJ}
\affil[2]{iMolecule, Skolkovo Institute of Science and Technology, Moscow, Russia}
\affil[3]{Department of Chemistry, Rutgers University, Camden, NJ}

\date{}
\twocolumn
\begin{document}

\maketitle

\begin{abstract}
Predicting the physical interaction of proteins is a cornerstone problem in computational biology. New classes of learning-based algorithms are actively being developed, and are typically trained end-to-end on protein complex structures extracted from the Protein Data Bank. These training datasets tend to be large and difficult to use for prototyping and, unlike image or natural language datasets, they are not easily interpretable by non-experts.
We present Dock2D-IP and Dock2D-IF, two ``toy'' datasets that can be used to select algorithms predicting protein-protein interactions---or any other type of molecular interactions. Using two-dimensional shapes as input, each example from Dock2D-IP (``interaction pose'') describes the interaction pose of two shapes known to interact and each example from Dock2D-IF (``interaction fact'') describes whether two shapes form a stable complex or not.
We propose a number of baseline solutions to the problem and show that the same underlying energy function can be learned either by solving the interaction pose task (formulated as an energy-minimization ``docking'' problem) or the fact-of-interaction task (formulated as a binding free energy estimation problem).
\end{abstract}

\section{Introduction}

One of the long-term goals of computational structural biology is to explain and predict the phenotype of an organism from the basic physical and chemical properties of its molecular constituents. Given the key role of proteins in living processes, a central problem in the field is to predict the structure of a protein from its amino acid sequence~\cite{dill2012protein, alquraishi2021machine}. Considerable progress on that problem has recently been made by algorithms such as AlphaFold2~\cite{jumper2021highly} and RoseTTAFold~\cite{baek2021accurate}, using multiple sequence alignments (MSAs) as additional inputs, and by more recent models replacing MSAs with sequence embeddings from large language models~\cite{lin2022language,chowdhury2022single,weissenow2022protein, fang2022helixfold}. The outstanding performance of AlphaFold2 at the 2020 CASP~14 competition suggests that such architectures can solve a large portion of the protein structure prediction problem~\cite{callaway2020willchange, alquraishi2020watershed, akdel2022structural}.

Because proteins typically perform their functions by coming into contact with other proteins (as well as with nucleic acids, carbohydrates, and a variety of small molecules), it is also critical to be able to predict protein-protein interactions (PPIs) and the structures of the complexes they form. While this problem can be viewed simply as a more general version of the protein structure prediction problem---involving two protein chains instead of one~\cite{AlphaFoldMultimer}---, many of the data sources used in state-of-the-art protein structure prediction algorithms are not as easy to leverage for the prediction of PPIs. There are fewer structures of protein complexes than of single proteins and, more importantly, the co-evolution signal between physically interacting proteins is weaker and harder to extract than that between interacting regions of a single protein~\cite{ovchinnikov2014robust, hopf2014sequence, cong2019coevolution}. For that reason, PPI prediction models are expected to benefit greatly from strong priors based on molecular structure and energy~\cite{nar_Zhang2013}.

In light of the success of AlphaFold2, it has become clear that learning-based algorithms will be crucial in solving many of the remaining key problems in computational structural biology.
From the last decade of progress in computer vision and natural language processing, we expect that the design of these new learning-based algorithms will be driven by the development of standard performance metrics and well-curated datasets.
One such example from the field of computer vision is the ImageNet dataset~\cite{ImageNet}, which enabled the development of the AlexNet~\cite{AlexNet} and ResNet~\cite{ResNet} architectures, still considered the gold standard for computer vision applications.
Many of these learning algorithms have risen to prominence by first showing outstanding performance on smaller, simpler datasets such as MNIST~\cite{MNIST}. These datasets allow for fast prototyping and are generally more interpretable than the large, complex datasets ultimately used for training.

In the case of PPI prediction, learning-based algorithms are typically trained on structures extracted from the Protein Data Bank~\cite{PDB} and tested on datasets such as the Docking Benchmark~\cite{dockingbenchv5Zlab} and the CASP/CAPRI targets~\cite{CASP,CAPRI}. However, these datasets are impractical for prototyping and are not easily interpretable by humans, which makes it harder to analyze the advantages and drawbacks of any new algorithm.

This paper presents Dock2D-IP and Dock2D-IF, two simple, small-scale datasets that can be used to select algorithms predicting protein-protein interactions. Our aim is to provide an equivalent of the MNIST dataset for the molecular recognition problem, formulated here as an energy-based shape recognition problem. While we are designing the Dock2D datasets with protein-protein interactions in mind, they are generic enough to apply to any type of receptor-ligand interactions.

Protein-protein interaction data fall into two broad categories~\cite{keskin2016ppi, petschnigg2011interactive}: 1) structural data obtained from X-ray crystallography or high-resolution electron microscopy and 2) fact-of-interaction data obtained using proteomics methods. In this work we use a single energy function to generate datasets for both categories, so that prediction models can be trained from either type of data. To make the problem more easily tractable and interpretable, we formulate it in two dimensions instead of three and use a simple shape-based interaction potential similar to the one proposed by Katchalski-Katzir et al.~\cite{katchalski1992molecular}.

The paper is divided as follows: the ``Datasets'' section describes the generating function and the Dock2D datasets, the ``Models'' section describes algorithms for the pose prediction problem and for the fact-of-interaction prediction problem, and the ``Results'' section reports the performance of the proposed models and discusses their relative merits and their potential for transfer learning.

\section{Datasets}

Learning-based problems usually assume that the training data are sampled from an unknown probability distribution and that a solution consists in deducing this distribution from the samples and using it to infer new data. We design the Dock2D datasets following this general principle.

We construct both the Dock2D-IP and Dock2D-IF datasets by first generating two pools of protein-like shapes: one for training/validation and one for testing. We generate these shapes by randomly placing points inside a circle of 40~pixels in diameter and by computing a concave hull using the Alpha Shape Toolbox~\cite{alphashape}. The concavity parameter $\alpha$ and number of points $n$ are sampled from a list of predefined values.
All shapes have approximately the same size but have a contour that is more or less intricate (depending on $n$) and more or less concave (depending on $\alpha$).

\begin{figure}
\includegraphics[width=\linewidth,trim={0 0.25cm 0 1cm},clip]
{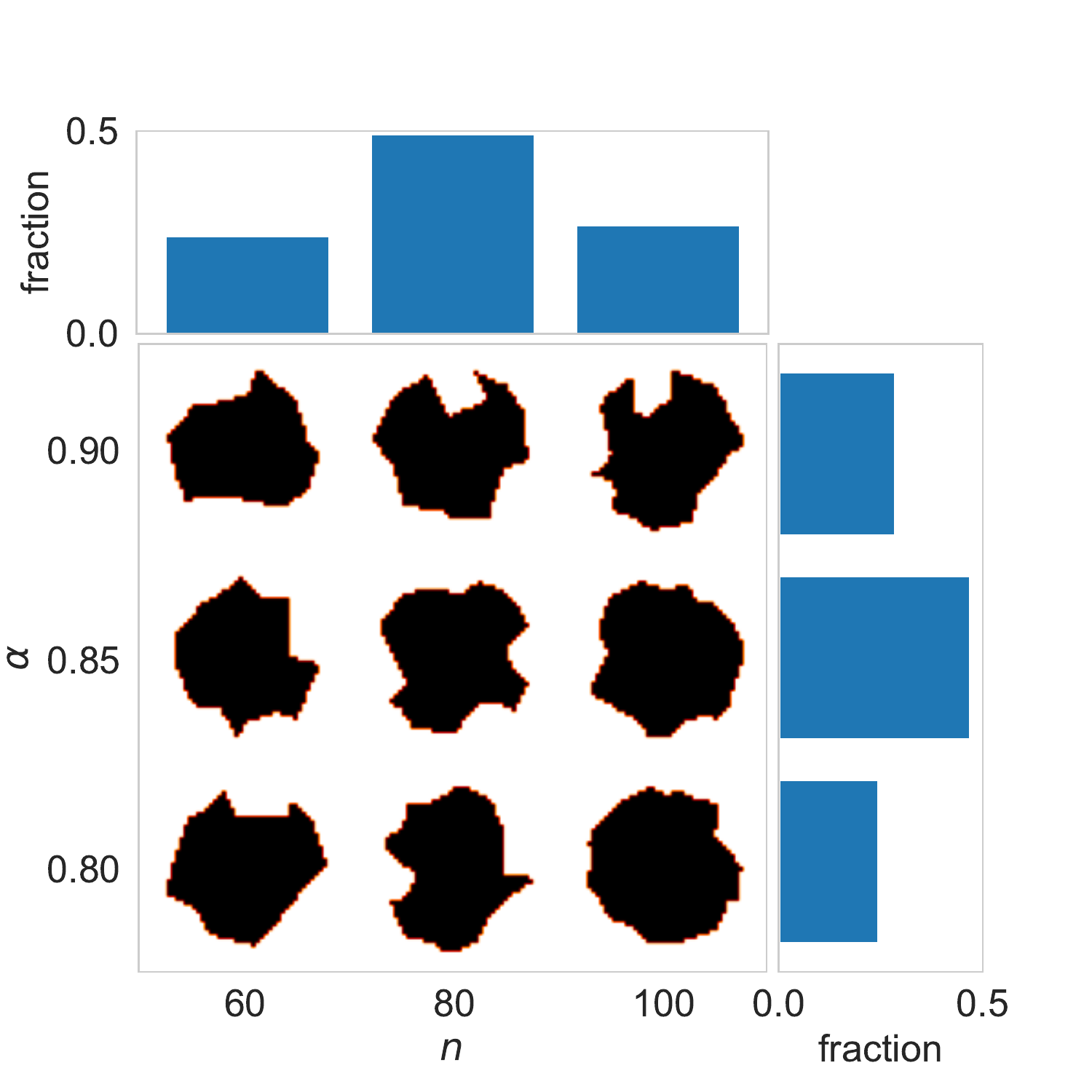}
\caption{Examples of training/validation shapes generated using predefined values of the $\alpha$ and $n$ parameters. The histograms show how often each value of $\alpha$ and $n$ was picked during shape generation. Both parameters are drawn from probability distributions (0.25, 0.50, 0.25).}
\label{fig:shapeDists}
\end{figure}

For the training/validation pool we pick $\alpha \in \{0.80, 0.85, 0.90\}$ and $n \in \{60, 80, 100\}$ (see Figure~\ref{fig:shapeDists}). In order to challenge the models, the test pool is created by picking $\alpha \in \{0.70, 0.80, 0.90, 0.95, 0.98\}$ and $n \in \{40, 60, 80, 100\}$ (see Figure~\ref{fig:test_shapeDist} for details). All shapes from each pool are then docked in a pairwise manner, effectively creating an ``interactome'' for which all combinations of ``receptor'' shape and ``ligand'' shape are assigned an optimal interaction pose and a binding free energy. Each pool consists of $N = 400$ shapes, for which $N(N+1)/2 = 80200$ distinct pairwise interactions are evaluated (400 homodimers and 79800 heterodimers).

We generate the interactome using an energy-based Boltzmann distribution
\begin{equation}
\label{eq:SampleDistr}
p(\mathbf{t}, \phi \,|\, R, L) = \frac{1}{Z}e^{-\beta E(\mathbf{t}, \phi, R, L)}
\end{equation}
where $R$ and $L$ represent the shapes of the receptor and the ligand, respectively, where $\mathbf{t} = (t_x, t_y)$ is the translation of the ligand with respect to the receptor, and where $\phi$ is the angle of rotation of the ligand around its center. Parameter $\beta$ stands for the inverse temperature and is set to 1 for convenience. Partition function $Z$ is defined such that $p$ sums up to 1 over the range of all possible translations and rotations.

\begin{figure*}[t]
\centering
\includegraphics[
width=\linewidth,% height=3cm,
trim={0cm 0cm 0cm 3cm},clip]
{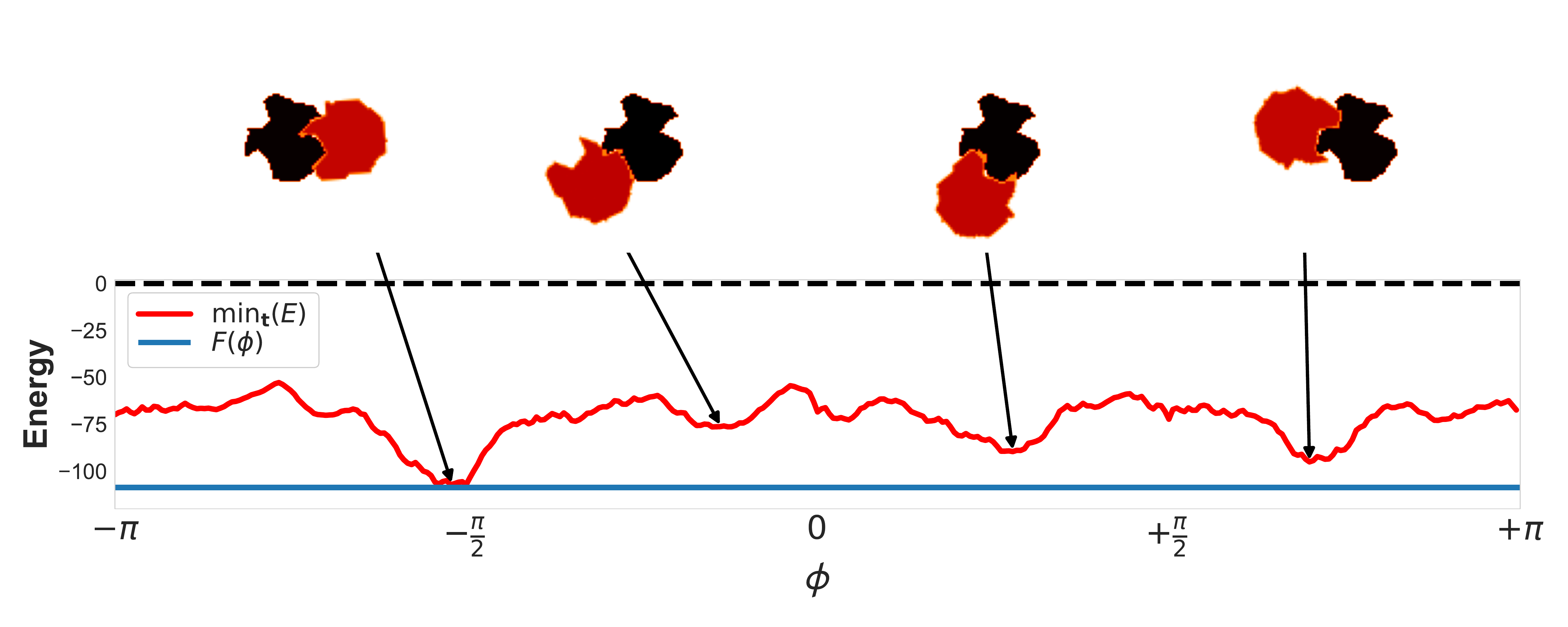}
\vspace{-0.75cm}
\caption{Docking energy of two shapes as a function of ligand orientation $\phi$ [see Eq.~(\ref{eq:Energy})]. Docked complexes are shown at four values of $\phi$, illustrating the global energy minimum (at $\phi_0 \sim -\frac{\pi}{2}$) and three local minima, with receptor shape in black and ligand shape in red.
The blue horizontal line represents the binding free energy $F = -\ln Z$ [see Eq.~\ref{eq:freeEnergy})], which is always less than the global energy minimum.}
\label{fig:pose_and_energysurface}
\end{figure*}

The energy function is defined as in Ref.~\cite{katchalski1992molecular}:
\begin{eqnarray}
\label{eq:Energy}
E(\mathbf{t}, \phi, R, L) &\!\!\!\!=\!\!\!\!& a_{00}\,\mathrm{corr}(\mathbf{t},  \phi, R, L) \nonumber\\
&& {}+a_{10}\,\mathrm{corr}(\mathbf{t}, \phi, R_\mathrm{s}, L) \nonumber\\
&& {}+a_{01}\,\mathrm{corr}(\mathbf{t}, \phi, R, L_\mathrm{s}) \nonumber\\
&& {}+a_{11}\,\mathrm{corr}(\mathbf{t}, \phi, R_\mathrm{s}, L_\mathrm{s})
\end{eqnarray}
where
\begin{equation}
\label{eq:Corr}
\mathrm{corr}(\mathbf{t}, \phi, R, L) = \int R(\mathbf{r}) M_\phi L(\mathbf{r}-\mathbf{t}) d\mathbf{r}
\end{equation}
The integral represents a summation over pixelated versions of the $R$ and $L$ shapes, and $M_\phi$ corresponds to a rotation of the ligand by angle $\phi$ around its center.
In Eq.~(\ref{eq:Energy}), $R$ and $L$ represent the bulk of the shape (1 if the center of the pixel is inside, 0 if it is outside), and $R_\mathrm{s}$ and $L_\mathrm{s}$ represent its boundary, obtained by convolving the image with a vertical and horizontal Sobel-Feldman $3 \times 3$ edge-detection filter~\cite{sobel-feldman}. %(The subscript ``s'' is for ``surface''.)
We set the coupling coefficients to the following values: $a_{00} = 100$, $a_{10} = a_{01} = -10$, and $a_{11} = -10$. The energy of a receptor-ligand conformation is low when the overlap of the $R$ and $L$ bulk regions is small while the overlap of the bulk of one shape with the boundary of the other is large (see Figure~\ref{fig:pose_and_energysurface}). The additional negative contribution of the boundary-boundary overlap penalizes the formation of non-specific interfaces.

We construct the Dock2D-IP dataset (``IP'' for ``interaction pose'') by minimizing the energy function (\ref{eq:Energy}) and selecting the transformation $(\mathbf{t}_0, \phi_0)$ that leads to that global energy minimum:
\begin{equation}
\label{eq:SampleMin2}
(\mathbf{t}_0, \phi_0) = \argmin_{\mathbf{t}, \phi} E(\mathbf{t}, \phi, R, L)
\end{equation}
The minimum energy itself is denoted $E_0$.
This is equivalent to letting $\beta \rightarrow \infty$, in which case the Boltzmann distribution would be nonzero for that docking pose only:
\begin{equation}
\label{eq:SampleMin}
p(\mathbf{t}, \phi \,|\, R,L) = \delta(\mathbf{t} - \mathbf{t}_0) \delta(\phi - \phi_0) \\
\end{equation}
Because it is created from minimum-energy poses, Dock2D-IP is also reminiscent of real training datasets for protein docking, which are usually composed of high-resolution X-ray structures representing a single conformation, not a Boltzmann-weighted ensemble.

The Dock2D-IP dataset contains all $(R,L)$ pairs for which $E_0 < -100$. This energy cutoff ensures that all examples have at least one strong binding mode.
Although proteins displaying multiple binding modes are generally selected against during natural evolution (because they often lead to aggregation), we do not explicitly discard examples having multiple interaction poses with energies below the cutoff.

The Dock2D-IP training/validation set contains 5081 examples, which are shuffled and split 4:1, resulting in 4064 training examples and 1017 validation examples. (Note that any shape found with a certain interaction partner in a training example may be found with a different partner in a validation example. For validation purposes, those two examples are considered different enough.) The test set contains 11850 examples. Each Dock2D-IP example is stored as two $50 \times 50$ images $R$ and $L$, along with the transformation $(\mathbf{t}_0, \phi_0)$ that brings $L$ in its optimal pose relative to $R$.

Because the shapes are pixelated before they are rotated, the energy tends to be lower when $\phi$ is a multiple of $90^\circ$. For this reason, values $\phi_0 = -180$, $-90$, $0$, and $90^\circ$ are overrepresented in the dataset (see Figure~\ref{fig:sup_rotation_dists} in Appendix). We do not correct for that slight imbalance.

We construct the Dock2D-IF dataset (``IF'' for ``interaction fact'') by computing the following free energy of interaction:
\begin{equation}
\label{eq:freeEnergy}
F(R,L) = -\ln Z = -\ln \sum_{\mathbf{t}, \phi} e^{-E(\mathbf{t}, \phi, R, L)}
\end{equation}
The sum over $\mathbf{t}$ is performed on a $100\times 100$ grid, representing translations from $-50$ to $+50$~pixels in each direction, while the sum over $\phi$ is performed from $-180$ to $179^\circ$ in $1^\circ$ increments.
We consider two shapes $R$ and $L$ to interact if $F(R,L) < -100$, assigning a ground-truth label ``1'' for shapes that interact and ``0'' for shapes that do not. The $F < -100$ condition ensures that those shapes have at least one binding mode with $E_0 < -84.90$, which corresponds to $-100 + \ln(360\times 100\times 100)$ (see Figures~\ref{fig:pose_and_energysurface} and \ref{fig:sup_energy_dists}). 
This derives from the fact that free energy $F$ can be no higher than the minimum energy $E_0$ (if only one state of the system has $E = E_0$ and all other states have energies $E \rightarrow +\infty$) and can be no lower than $E_0 - \ln(n)$ (if all $n$ states of the system have the same energy $E_0$).

\begin{figure}[t]
\includegraphics
[
width=1.0\linewidth,
% height=2in,
% width=3in,
% trim={14cm 6cm 14cm 2cm},clip
trim={8.5cm 5cm 6cm 2cm},clip
]
{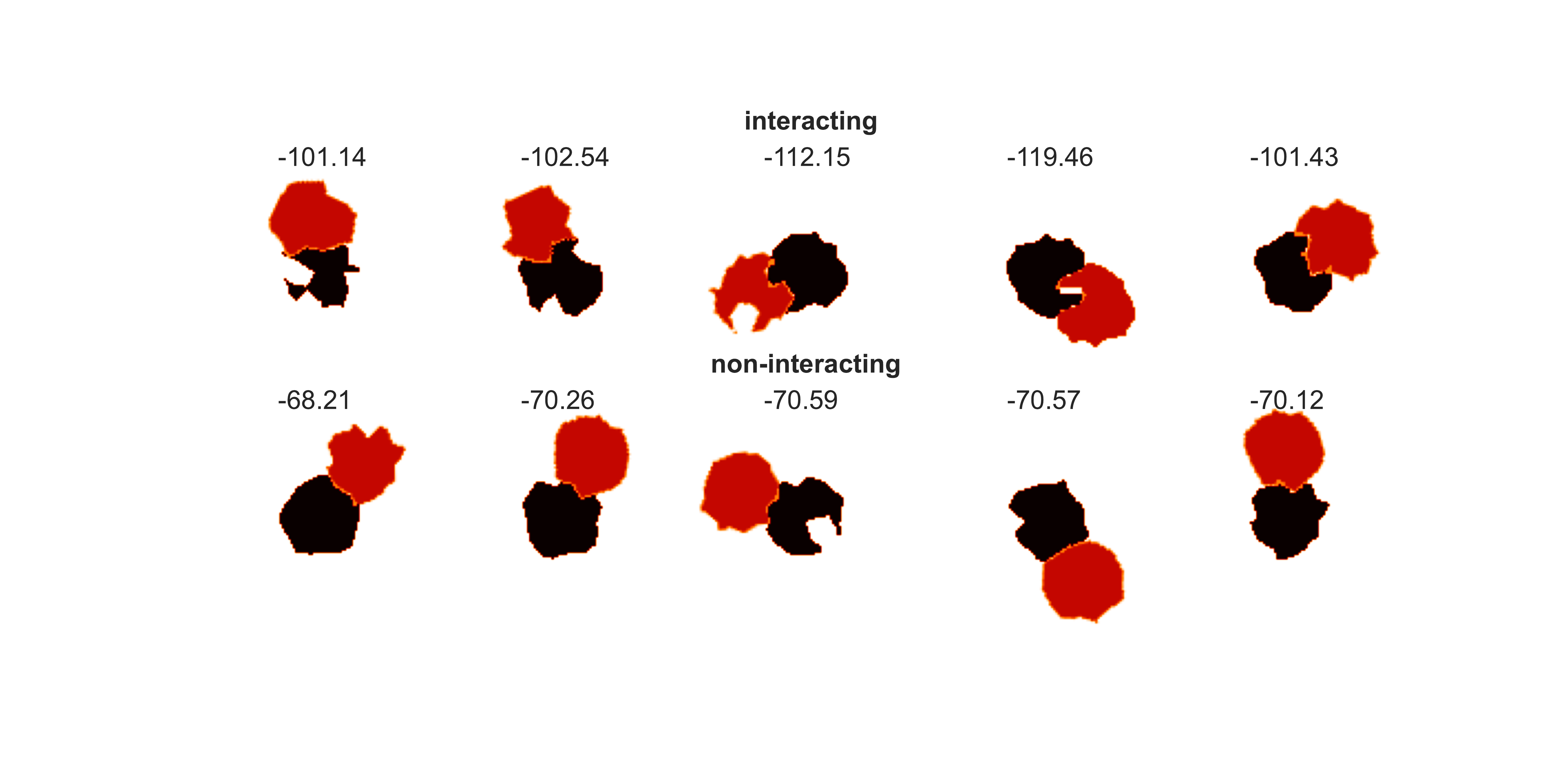}
\caption{Examples of conformations of shapes from the training/evaluation set forming a positive interaction (top row) and a negative interaction (bottom row). The number above each conformation is the binding free energy $F$.}
\label{fig:interacting_noninteracting_examples}
\end{figure}

The Dock2D-IF training/validation dataset includes all 80200 pairwise interactions from the pool of 400 shapes and is split 4:1, resulting in 64160 training examples and 16040 validation examples. With our choice of free energy threshold, approximately 7\% of all pairs have positive interactions: 4481/64160 (7.0\%) for the training examples and 1170/16040 (7.3\%) for the validation examples. The test set contains 80200 examples, of which 13025 (16.2\%) have positive interactions. Figure~\ref{fig:interacting_noninteracting_examples} shows examples of binding modes for interacting and non-interacting examples from the dataset (see also Figure~\ref{fig:interacting_noninteracting_examples_test} in Appendix).

Interestingly, the Dock2D datasets show a high propensity for homodimers. In the Dock2D-IP dataset, 143 of the 5081 training/validation examples and 178 of the 11850 test examples are homodimers. This represents 2.8\% and 1.5\%, respectively, while we would have expected a ratio of 400/80200 = 0.5\% if the probability of binding had been the same for any two shapes.
The enrichment in homodimers is even more pronounced in the Dock2D-IF training/validation set, where 35.8\% of the 400 homodimeric interactions are positive (143/400) while only 6.9\% of the 79800 heterodimeric interactions are positive (5508/79800). In the Dock2D-IF test set, 45.0\% of homodimeric interactions are positive (180/400) and 16.1\% of heterodimeric interactions are positive (12845/79800).
This is a known effect in proteins \cite{book_Klotz75, arbbs_Goodsell2000}, which has been attributed to a number of biophysical or evolutionary reasons~\cite{pnas_Andre2008}. In the Dock2D datasets, the reason is purely geometric and follows the argument originally proposed by Lukatsky and coworkers~\cite{prl_Lukatsky2006, jmb_Lukatsky2007}. Since homodimers have a twofold ($C_2$) symmetry, any region of a homodimer interface contributes to the stabilization energy twice: once on each side of the symmetry axis. Therefore, it is more likely that shapes generated at random will form strong interactions with copies of themselves (homodimers) than with unrelated shapes (heterodimers).
Homodimer interactions are also artificially enhanced by the fact that they correspond to $\phi_0 = -180^\circ$, which is a multiple of $90^\circ$ (see Figure~\ref{fig:sup_rotation_dists} in Appendix).

\section{Models}

As stated in the Introduction, the main goal of the Dock2D datasets is to facilitate the development of models that can learn from either ``interaction pose'' (IP) data or ``interaction fact'' (IF) data. Here we propose two types of models that set a standard of algorithmic performance on each dataset: a ``brute-force'' model, that generates a prediction by calculating the energy function for all translations and rotations, and an ``economical'' model, that learns the energy function without calculating the full partition function $Z$.
While the ``brute-force'' models essentially solve both the IP and IF prediction tasks, they are impractical solutions for the three-dimensional version of the problem, for which the number of states is much larger than in the two-dimensional version. By contrast, the ``economical'' models aim to provide tractable solutions in higher dimensions.

\subsection{Overall architecture}

\begin{figure*}[t]
\begin{center}
\includegraphics[scale=0.45]{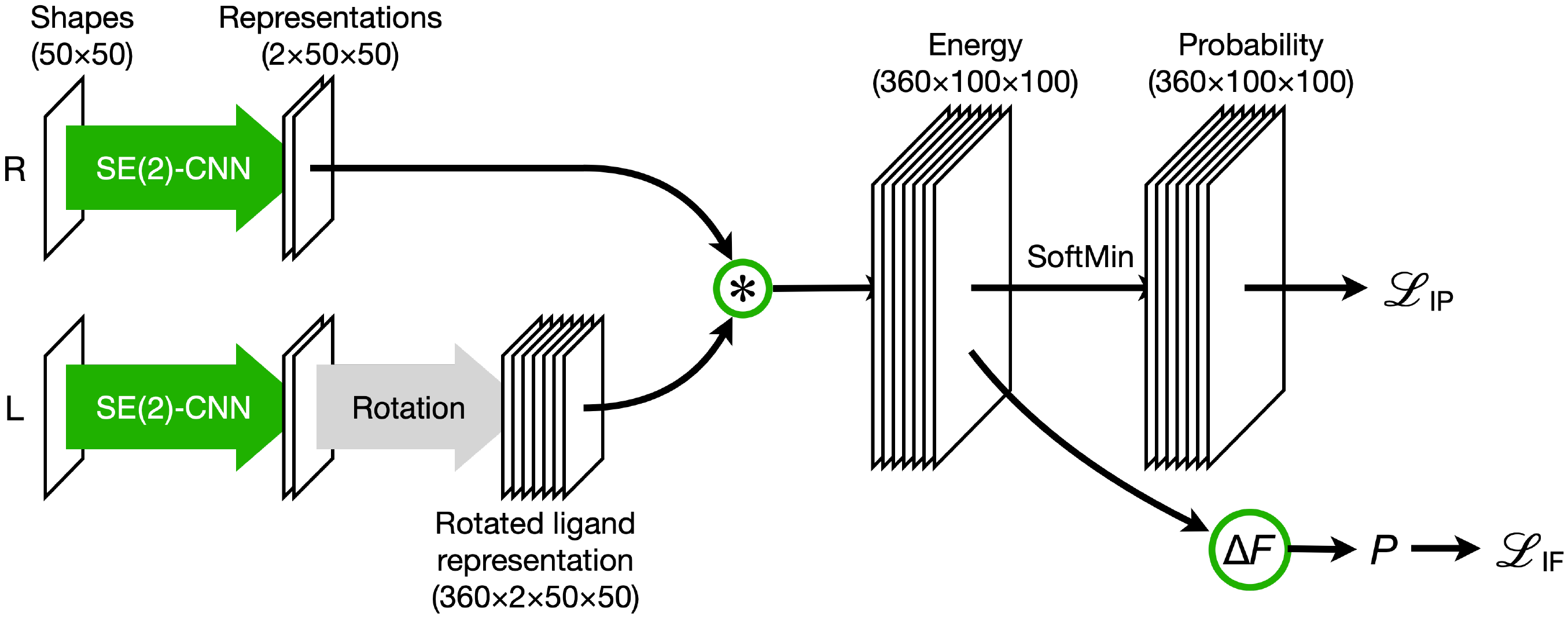}
\end{center}
\vspace{-1.5em}
\caption{Brute-force solution to the joint ``interaction pose'' (IP) and ``interaction fact'' (IF) problems. The receptor and ligand shapes are converted to their two-component representations, and rotated ligand representations are explicitly generated for $\phi = -180^\circ$ to $+179^\circ$, in increments of $1^\circ$.
The asterisk ($\boldsymbol{\ast}$) represents the operation described by Eq.~(\ref{eq:Energy}), leading to an energy $E(\mathbf{t},\phi)$ and a probability $P(\mathbf{t},\phi)$ from which $\mathcal{L}_\mathrm{IP}$, the loss function for the IP prediction task, is directly computed. 
$\Delta F$ is computed using Eq.~(\ref{eq:deltaF}), leading to a probability of binding $P$ from which $\mathcal{L}_\mathrm{IF}$ is directly computed.
Operations shown in green contain learnable parameters. The architecture is similar for the ``simplified IP task'', with a single angular value calculated (the ground-truth angle) instead of 360. For the ``sampled IF task'', the 360 angular values are replaced by $n(B) < 360$ values sampled using a Monte Carlo approach (see text).}
\label{fig:DockerModel}
\end{figure*}

All models learn a two-feature representation of each shape (receptor and ligand) using a Siamese SE(2)-equivariant CNN~\cite{e2cnn} (see Figure~\ref{fig:DockerModel}) and use these features to compute an energy of the form~(\ref{eq:Energy}), with weights $a_{00}$, $a_{10}$, $a_{01}$, and $a_{11}$ treated as learnable parameters. Spatial correlations~(\ref{eq:Corr}) are computed using fast Fourier transform (FFT)~\cite{pytorch}.

The Siamese SE(2)-equivariant CNN module has only two convolutional layers, each using a fixed kernel size of $5\times5$. The first layer maps the input shape onto one scalar and four vectors and the second layer maps those back to one scalar and one vector, representing the bulk and boundary features used in Eq.~(\ref{eq:Energy}). Each convolutional layer is followed by a ReLU applied to the norm of the features~\cite{e2cnn}, to preserve the equivariance and allow the scalar and the vectors to mix. The final vector feature is reduced to scalar by taking an element-wise norm.

\subsection{Docking models}

For the interaction pose (IP) prediction task, the training loss $\cal{L}_\mathrm{IP}$ is defined as the cross-entropy between the predicted Boltzmann distribution
\begin{equation}
\label{eq:boltzmann_dist}
P(\mathbf{t}, \phi) = \frac{ e^{-E(\mathbf{t}, \phi)}}{\sum_{\mathbf{t}, \phi} e^{-E(\mathbf{t}, \phi)}}
\end{equation}
and the ground-truth probability distribution, encoded as $P = 1$ at $(\mathbf{t}_0, \phi_0)$ and $P = 0$ everywhere else. The IP task requires to learn all parameters of the CNN, as well as the four ``$a$'' parameters from Eq.~(\ref{eq:Energy}).

This algorithm is called ``brute-force'' because it explicitly computes the energy for all possible translations and rotations. While a brute-force approach is entirely feasible in two dimensions, it becomes somewhat prohibitive in three dimensions, requiring a six-dimensional array of energy values to be computed. Nevertheless, brute-force protein docking algorithms are commonly used~\cite{huang2014search}.

A simpler version of the IP task is also devised, in which the correct orientation $\phi_0$ is provided and the model is trained to only predict the correct translation $\mathbf{t}_0$. The training loss is defined as previously but using the predicted Boltzmann distribution over the translations only:
\begin{equation}
\label{eq:boltzmann_dist_simplified}
P(\mathbf{t}, \phi_0) = \frac{ e^{-E(\mathbf{t}, \phi_0)}}{\sum_{\mathbf{t}} e^{-E(\mathbf{t}, \phi_0)}}
\end{equation}
This ``simplified IP task'' can in principle learn the same parameters as the full IP task.

\subsection{Binding free energy models}

For the interaction fact (IF) prediction task, energy $E(\mathbf{t}, \phi)$ is further transformed into a single value $\Delta F$ meant to represent the binding free energy of the two shapes.
The training loss $\cal{L}_\mathrm{IF}$ is defined as the binary cross-entropy between the probability of binding
\begin{equation}
P = \frac{e^{-\Delta F}}{e^{-\Delta F} + 1}
\label{eq:prob_binding}
\end{equation}
and the ground-truth fact of interaction ($P=1$ or $P=0$).
The free energy of binding is written as
\begin{equation}
\Delta F = -\ln Z - F_0
\label{eq:deltaF}
\end{equation}
In the ``brute-force'' version of the IF model, the partition function $Z$ is evaluated by integrating over all translations and rotations:
\begin{equation}
\label{eq:brute_force_z}
Z = \sum_{\mathbf{t},\phi} e^{-E(\mathbf{t}, \phi)}
= \sum_\phi e^{-F(\phi)}
\end{equation}
with $e^{-F(\phi)} = \sum_{\mathbf{t}} e^{-E(\mathbf{t}, \phi)}$. Constant $F_0$ is a learnable parameter that represents the free energy level below which the shapes are considered to interact (when $-\ln Z < F_0$, $\Delta F < 0$ and $P > 0.5$). The IF task requires to learn all parameters from the IP task, plus the $F_0$ parameter.

Equations~(\ref{eq:prob_binding}) and (\ref{eq:deltaF}) involve exponential functions (``sigmoid'' and ``logsumexp'', respectively), which can have their gradients vanish if any of their arguments gets out of bounds. To avoid those situations, we add a regularization loss that prevents any value of $\Delta F$ from becoming too large:
\begin{equation}
{\cal L}_\mathrm{reg} = w |\Delta F|
\end{equation}
Weight $w$ is chosen to be small, since the only purpose of this loss term is to preserve a nonzero gradient.

A second version of the IF model, called ``sampled'', learns the same energy function as the brute-force model but using only a subset of all possible orientations. Instead of computing the partition function from Eq.~(\ref{eq:brute_force_z}), we compute the following estimate of it:
\begin{equation}
\label{eq:sampled_z}
\hat Z = \sum_{\phi \in B} e^{-F(\phi)} + \sum_{\phi \notin B} e^{-F_0'}
\end{equation}
$F(\phi)$ is computed explicitly for orientations $\phi \in B$ but, for orientations $\phi \notin B$, it is assumed to have a uniform value $F_0'$, treated as an additional learnable parameter. Estimate $\hat Z$ is expected to provide the same training signal as $Z$ as long as the angles $\phi \in B$ correspond to the dominant interactions between the shapes. $\hat Z$ becomes equal to $Z$ when $B$ includes all angles $\phi$ used in the brute-force calculation of $Z$, that is, all integers from $-180$ to $+179$.

Sample $B$ is defined separately for each training example and is generated using a Monte Carlo procedure.
After assigning each example a random initial orientation $\phi$, a new orientation $\phi'$ is picked by performing 100 steps of random walk (changing the orientation by $\pm 1^\circ$ at each step). Following the conventional Metropolis algorithm, the new value $\phi'$ is directly accepted if $F(\phi') \leq F(\phi)$ and is accepted with a probability $e^{-[F(\phi') - F(\phi)]}$ if $F(\phi') > F(\phi)$. Any accepted value is added to $B$, unless it is already in. Two Monte Carlo moves are attempted each time the training example is presented to the network, which may add up to 2 new elements to $B$.
Since no element are ever removed from $B$ during training, $\hat Z$ could eventually become equal to $Z$ for all examples, at which point parameter $F_0'$ would be effectively removed from the model and would become unlearnable. 
However, this point is never reached in practice. First, since each Monte Carlo move consists in 100 steps of random walk on a ring, the sampling procedure can visit only half of all 360 $\phi$ values: either the odd values or the even values. Second, even after 1000 epochs of training, $B$ samples on average 32.8\% of all 180 possible orientations per training example (see Figure~\ref{fig:rotation_surface_saturation} in Appendix).

\section{Results}

\begin{table*}[h]
\caption{Performance (mean ligand RMSD) of models on the validation and test sets of the interaction pose dataset (Dock2D-IP). The entire validation and test sets were used in evaluation (1017 and 11850 examples, respectively). The last two rows report how the features and energy function learned from the IF task perform on the IP task.}
\vspace{-0.4cm}
\begin{center}
\begin{tabular}{l|c|c|c}
\hline
Model & \# training examples & Validation & Test \\
\hline
Brute-force IP & 1000 & 1.04 & 1.54 \\
               &  100 & 1.01 & 1.39 \\
               &   10 & 4.24 & 6.11 \\
Simplified IP  & 1000 & 0.72 & 1.17 \\
               &  100 & 1.18 & 1.84 \\
               &   10 & 5.97 & 7.64 \\
\hline
Brute-force IF ${}^{a}$
               & 5050 & 0.53 & 0.82 \\
Sampled IF ${}^{b}$
               & 5050 & 1.27 & 2.05\\
\hline
\end{tabular}
\end{center}
\footnotesize{${}^{a}$ All learnable parameters transferred from ``brute-force IF'' model (and frozen).} \\
\footnotesize{${}^{b}$ All learnable parameters transferred from ``sampled IF'' model (and frozen).}
\label{tbl:IPResults}
\end{table*}

\begin{table*}[h]
\caption{Performance of the models on the validation and test sets (16040 and 80200 examples) of the fact-of-interaction dataset (Dock2D-IF). All models were trained on 100 pairs (5050 training examples). MCC is the Matthews correlation coefficient.} 
\vspace{-0.4cm}
\begin{center}
\begin{tabular}{l|l|c|c|c|c}
\hline
Model & Subset & Accuracy & Precision & Recall & MCC \\
\hline
Brute-force IF
& Validation & 0.99 & 0.98 & 0.93 & 0.95 \\
& Test & 0.99  & 0.99 & 0.93 & 0.95 \\
%\hline
Sampled IF 
& Validation & 0.99 & 0.91 & 0.92 & 0.91 \\
& Test & 0.97 & 0.90 & 0.94 & 0.90 \\
\hline
Brute-force IF$_\mathrm{IP}$ ${}^{a}$
& Validation & 0.99 & 0.98 & 0.92 & 0.95 \\
& Test & 0.98 & 0.99 & 0.91 & 0.95 \\
%\hline
Brute-force IF$_\mathrm{IP}$ ${}^{b}$
& Validation & 0.99 & 0.97 & 0.94 & 0.95 \\
& Test & 0.98 & 0.97 & 0.93 & 0.94 \\
\hline
\end{tabular}
\end{center}
\footnotesize{${}^{a}$ SE(2)-CNN weights transferred from ``brute-force IP'' model (and frozen). Only $a_{00}$, $a_{01}$, $a_{10}$, $a_{11}$, and $F_{0}$ are learned.} \\
\footnotesize{${}^{b}$ SE(2)-CNN weights transferred from ``simplified IP'' model (and frozen). Only $a_{00}$, $a_{01}$, $a_{10}$, $a_{11}$, and $F_{0}$ are learned.}
\label{tbl:IFResults}
\end{table*}

Table~\ref{tbl:IPResults} shows the performance of the IP models on the Dock2D-IP validation and test sets. Although 4064 training examples are available in the Dock2D-IP dataset, the models were trained on fewer examples (1000, 100, and 10) to illustrate their data efficiency. All IP models were trained for 100 epochs (see Figure~\ref{fig:sup_IP_training_curves} for the training curves).
Models reported in Table~\ref{tbl:IPResults} have mean ligand RMSDs below 2 pixels, indicating that they recover the true poses within the resolution limits of the input.
The learned representations, shown in Figure~\ref{fig:input_and_feat_shapes}, indicate that the models have also recovered the equivalent of the bulk and boundary features used by the generating function.

Interestingly, the performance of the model trained on the ``simplified'' IP task is comparable to that of the model trained on the full IP task (see Table~\ref{tbl:IPResults}).
We also note that very few training examples are needed to achieve optimal performance: Using 100 examples instead of 1000 does not significantly affect the performance. The performance is somewhat reduced when only 10 examples are used but that degradation is due to a small fraction of the validation/test examples for which an alternate pose is predicted ($\mathrm{RSMD} > 10$) (see Figure~\ref{fig:RMSD_histogram_BFIP_10ex}).

Table~\ref{tbl:IFResults} shows the performance of the IF models on the Dock2D-IF validation and test sets (see Figure~\ref{fig:sup_IF_training_curves} for the training curves). Interpretation is complicated by the class imbalance of the dataset, which is about 15:1 for the training/validation set (15 non-interacting pairs for each interacting pair) and about 6:1 for the test set. To facilitate the comparison, we also report the performance using Matthews correlation coefficient (MCC)~\cite{chicco2020advantages}.
All brute-force IF models have high accuracy and precision, whether they are trained in full (``brute-force IF'') or using SE(2)-CNN weights transferred from the IP model (``brute-force IF$_\mathrm{IP}$''). The lower recall values reflect the 1:15 and 1:6 imbalance between positive and negative examples.
Moreover, each model performs similarly on both the validation and test sets, despite the difference is class imbalance and the greater variety of shapes in the test set.

The sampled IF model learns the energy function despite the free energy being poorly converged. After 1000 epochs of training, the Monte Carlo algorithm has typically visited 60 orientations per example (see Figure~\ref{fig:rotation_surface_saturation}), which means that $\hat Z$ from Eq.~(\ref{eq:sampled_z}) is computed using the default $F_0'$ value for typically 120 of the 360 discrete orientations.

The energy function learned from the IF task is directly transferable to the IP task and yields a comparable---if not better---performance (see the bottom two rows of Table~\ref{tbl:IPResults}).
By contrast, the energy function learned from the IP task, which has an undetermined scale because it only serves to identify the lowest-energy pose, can be transferred to the IF task only after appropriate re-scaling.
We perform that calibration by transferring to the IF model all parameters learned from the IP model but by re-training the ``$a$'' coefficients (see Eq.~(\ref{eq:Energy})).
The corresponding ``brute-force IF$_\mathrm{IP}$'' models, which are pre-trained on the IP task and have only five learnable parameters (the ``$a$'' coefficients and $F_0$; see Eq.~(\ref{eq:deltaF})), perform on par with the brute-force IF model (see the second half of Table~\ref{tbl:IFResults}).

The features learned from the IP task are very similar whether they are trained on the ``brute-force IP'' task or on the ``simplified IP'' task, but are somewhat different from those learned from the IF task (see Figure~\ref{fig:input_and_feat_shapes}). As expected, the IF task recovers the ground-truth features better, since it is directly transferable to the IP task and therefore more general.

\newcommand\w{1.7}
\newcommand\h{1.7}
\begin{figure}[!t]
\begin{centering}

\begin{tabular}{p{0.40\linewidth} p{0.55\linewidth}}
 \small A & \small B
\end{tabular}
\vspace*{-0.5mm}
\includegraphics[width=\w cm,height=\h cm]
{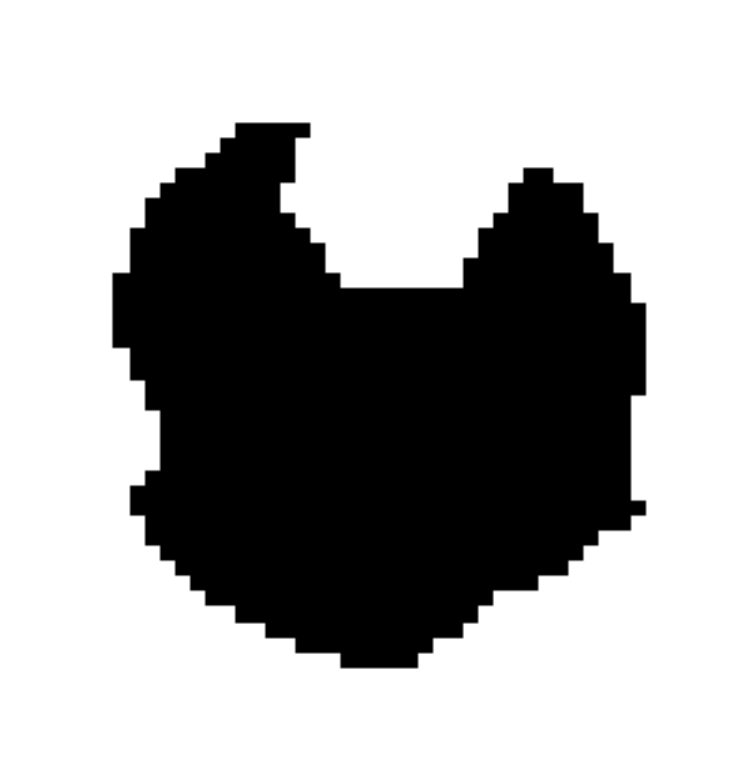}
\includegraphics[width=\w cm,height=\h cm]
{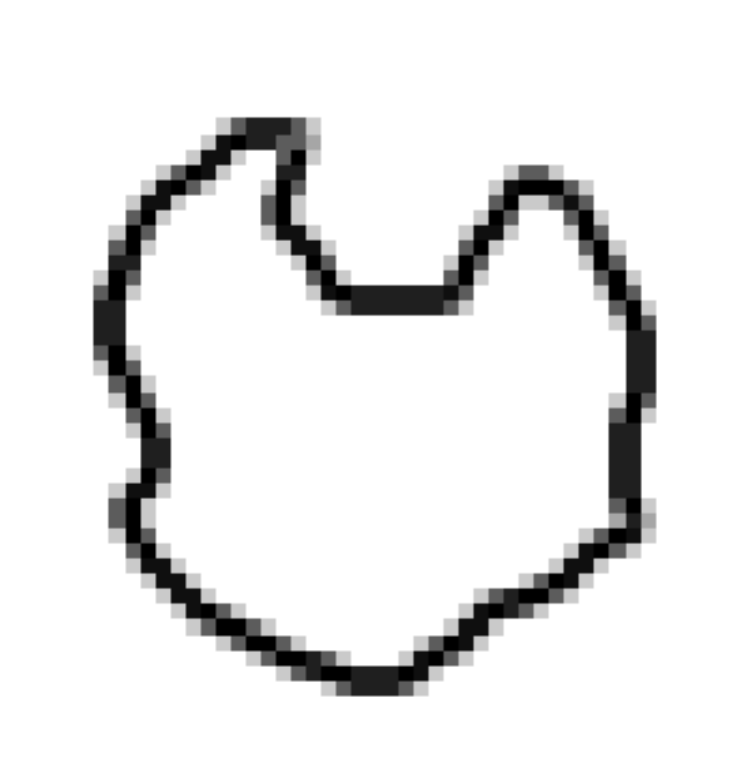}\
\includegraphics[width=\w cm,height=\h cm]
{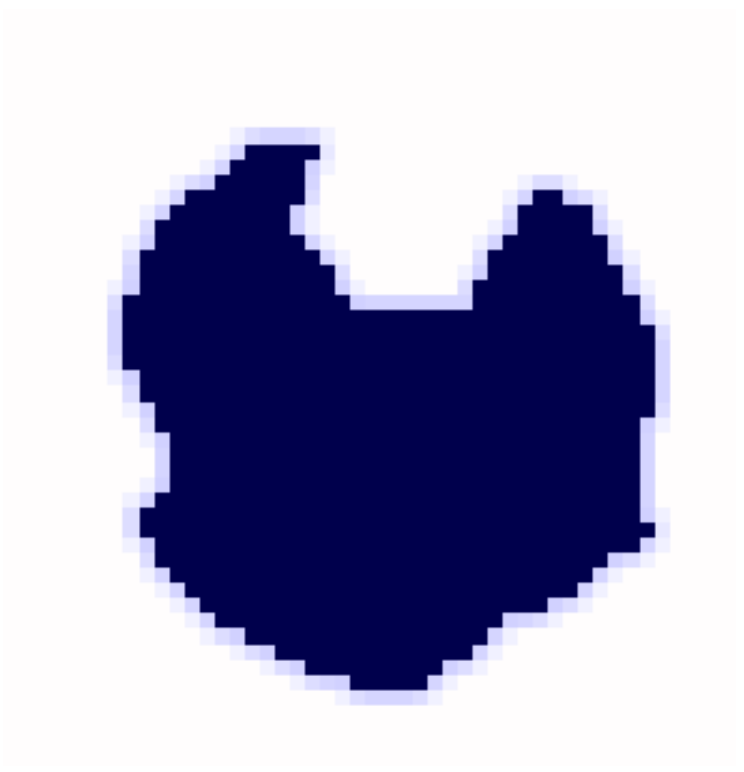}
\includegraphics[width=\w cm,height=\h cm]
{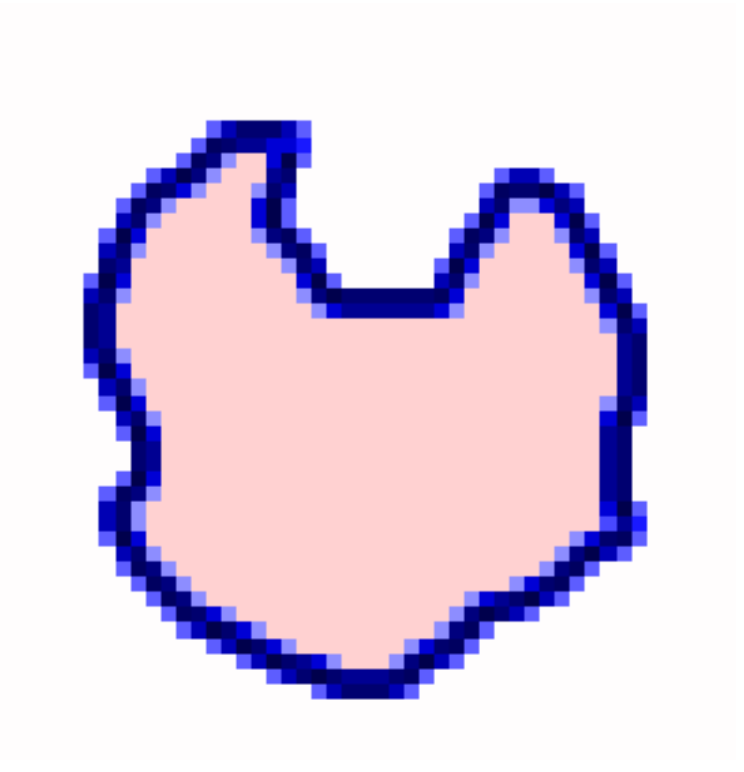}

\begin{tabular}{p{0.40\linewidth} p{0.55\linewidth}}
\small C & \small D
\end{tabular}
\vspace*{-0.5mm}
\includegraphics[width=\w cm,height=\h cm]
{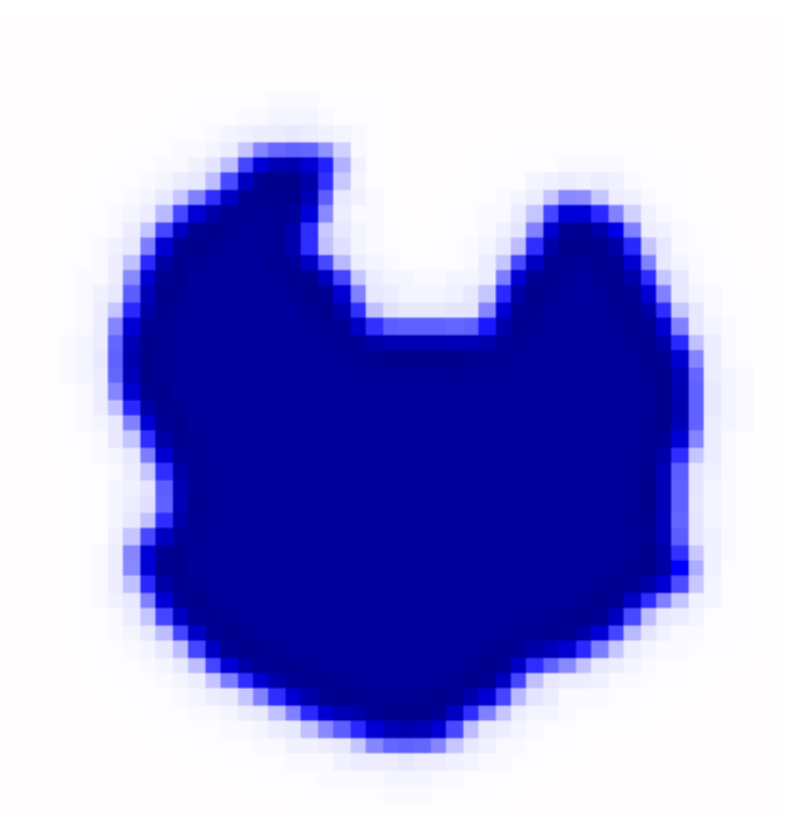}
\includegraphics[width=\w cm,height=\h cm]
{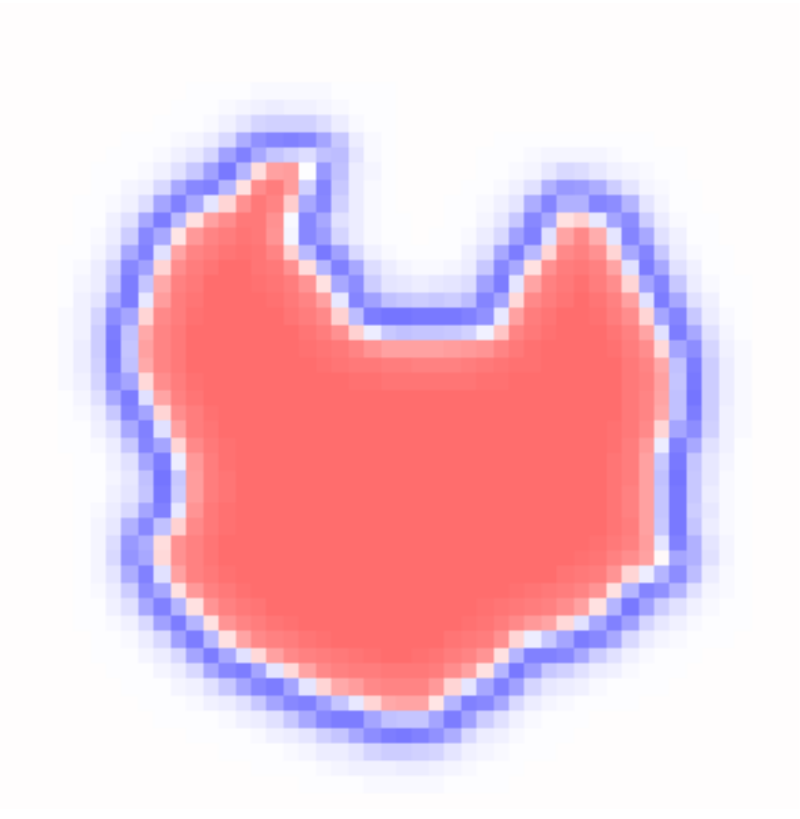}
\includegraphics[width=\w cm,height=\h cm]
{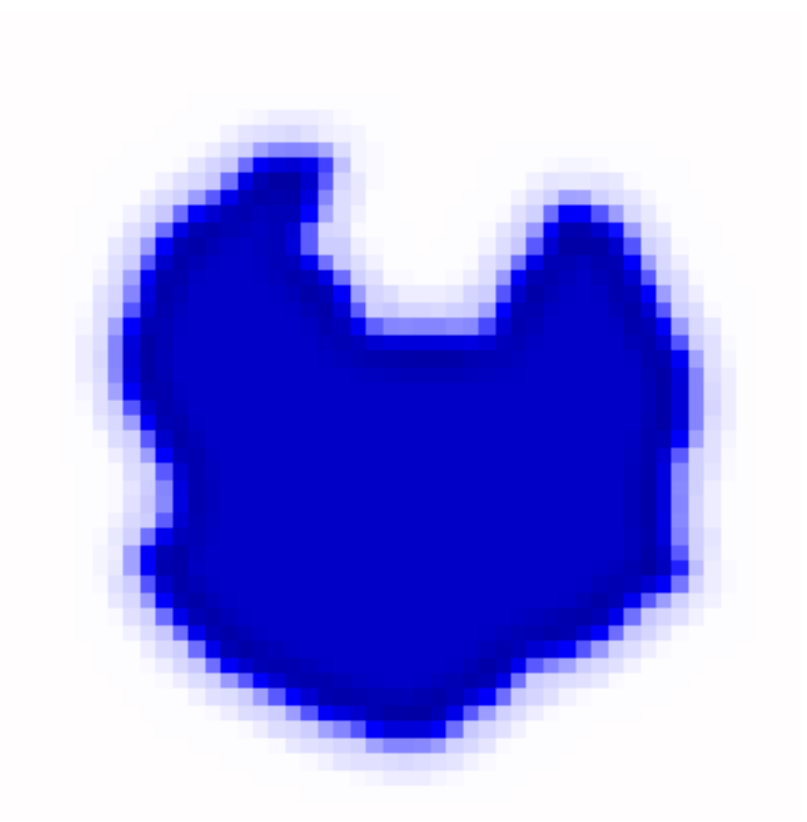}
\includegraphics[width=\w cm,height=\h cm]
{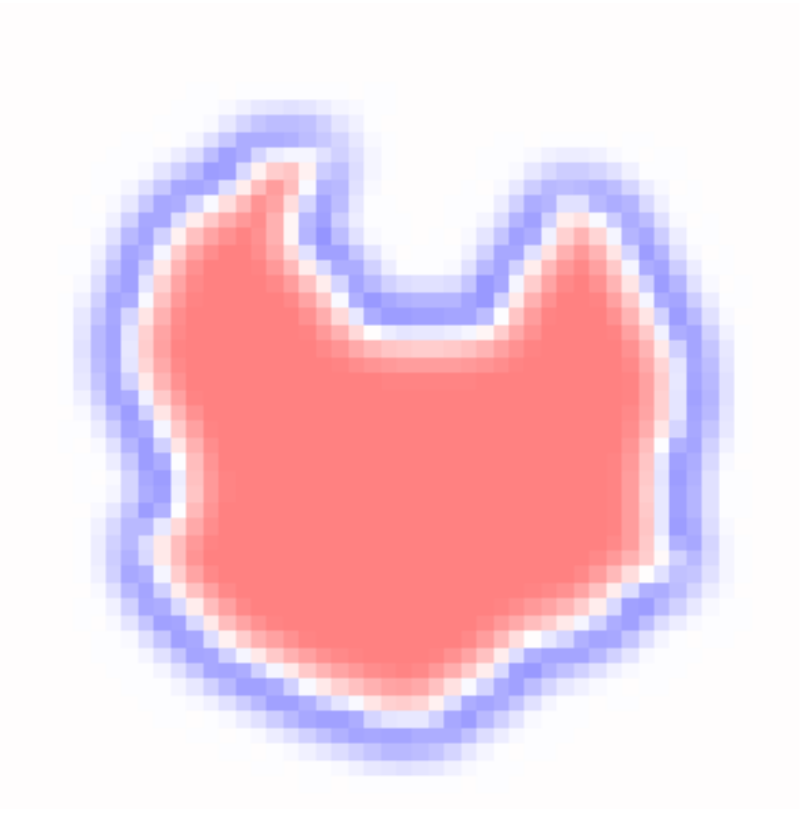}

\begin{tabular}{p{0.40\linewidth} p{0.55\linewidth}}
\small E & \small F
\end{tabular}
\includegraphics[width=\w cm,height=\h cm]
{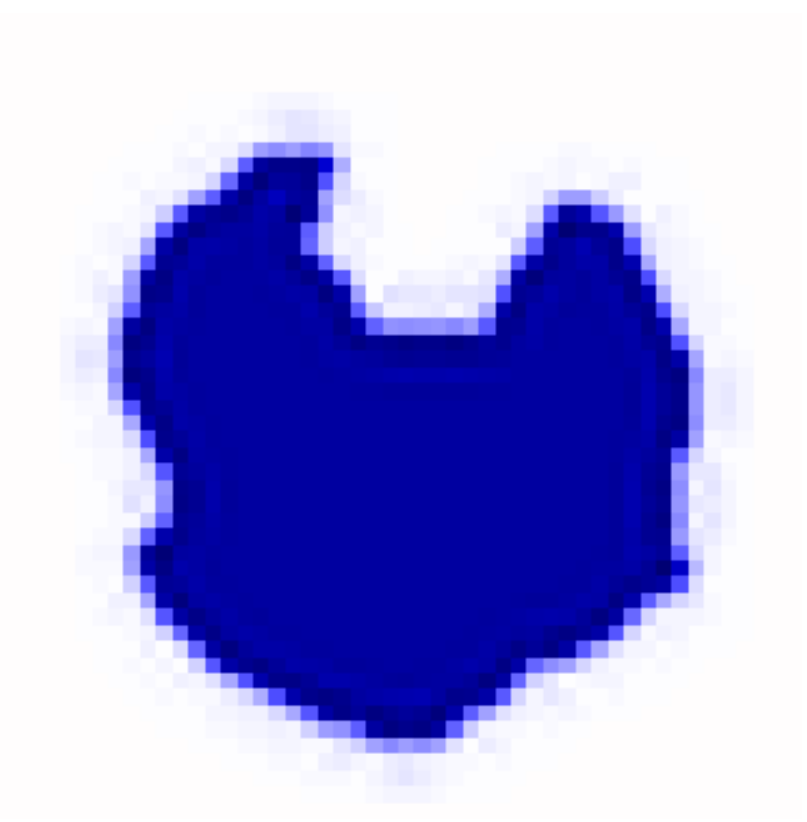}
\includegraphics[width=\w cm,height=\h cm]
{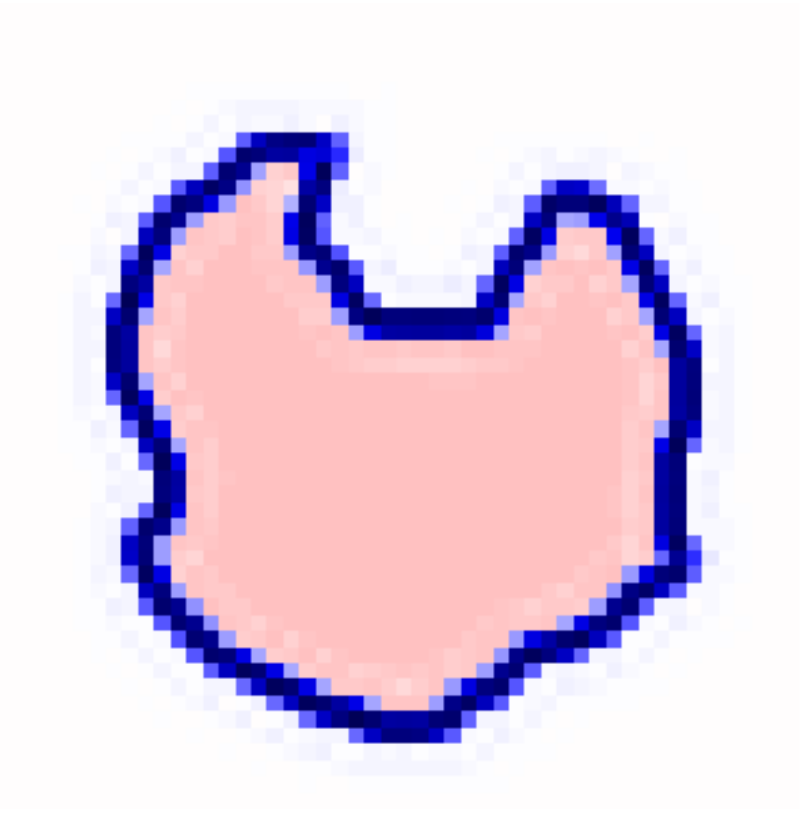}
\includegraphics[width=\w cm,height=\h cm]
{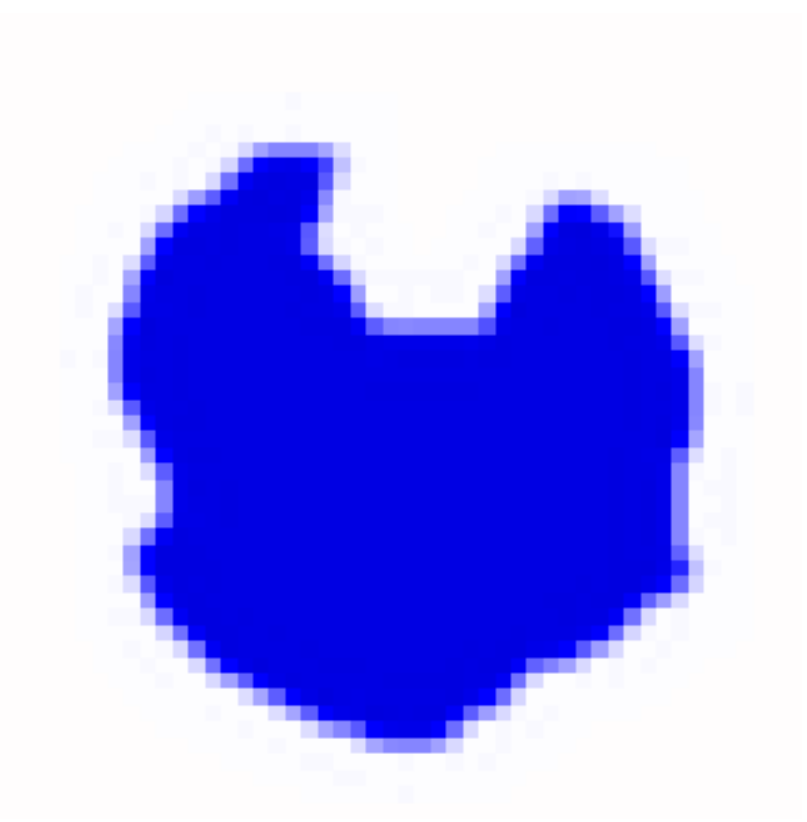}
\includegraphics[width=\w cm,height=\h cm]
{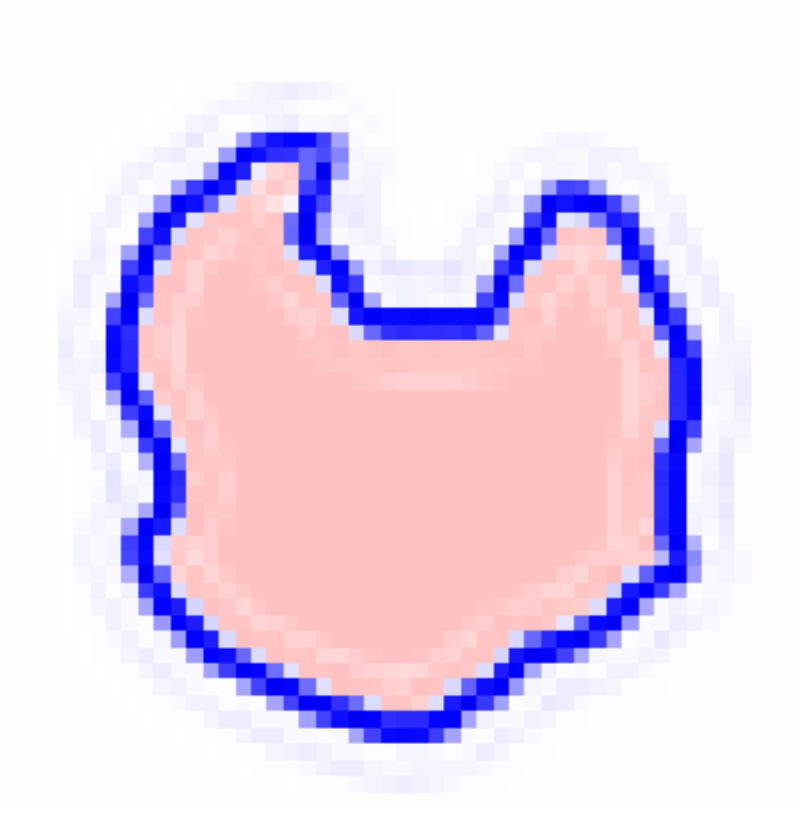}

\includegraphics[width=0.6\linewidth,height=0.8cm]
{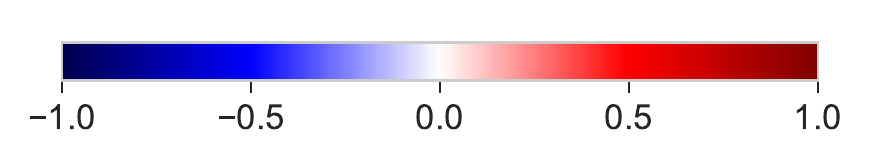}

\caption{Features learned from each model. Top row are the ground-truth features (panel A) and their orthogonalized versions (panel B). Second row are the orthogonalized features learned from the ``brute-force IP'' task (panel C) and ``simplified IP'' task (panel D). Third row are the orthogonalized features learned from the ``brute-force IF'' task (panel E) and ``sampled IF'' task (panel F).
Orthogonalization is based on the ground-truth ``$a$'' coefficients for the input features and on the learned ``$a$'' coefficients for the learned features.
}

\label{fig:input_and_feat_shapes}
\end{centering}
\end{figure}

\section{Conclusion}

We have proposed the Dock2D datasets (Dock2D-IP and Dock2D-IF) and several representation-learning ``baseline'' models demonstrating end-to-end learning of a simplified, 2D version of the molecular recognition problem. The representation-learning baselines effectively solved the interaction pose (IP) prediction problem in two dimensions (see Table~\ref{tbl:IPResults}), whether they were trained on the full task of predicting optimal pose $(\mathbf{t}_0,\phi_0)$ or on the simplified task of predicting optimal translation $\mathbf{t}_0$ given the ground-truth orientation $\phi_0$.
Training on the IP task (whether ``brute-force'' or ``simplified'') was extremely efficient, requiring only 100 epochs and completing in minutes.

The baselines also solved the fact-of-interaction (IF) prediction problem (see Table~\ref{tbl:IFResults}), whether the predicted $\Delta F$ values used in the $\cal{L}_\mathrm{IF}$ loss were computed explicitly using Eq.~(\ref{eq:brute_force_z}) or sampled using Eq.~(\ref{eq:sampled_z}).
While training models on the IF task took significantly longer---requiring 1000 epochs and sometimes taking days to complete---, training time can be drastically reduced by starting from the energy function learned from the IP task (see Figure~\ref{fig:sup_IF_training_curves}).

One of the goals of this work is to pave the way for algorithms using fact-of-interaction data to infer structural details of protein-protein complexes.
Our results suggest that, provided a model implements an energy function $E$ from which both an optimal pose and a binding free energy $F$ can be computed, the representations learned from training that model on the IP task are compatible with those learned from the IF task, and that a two-headed model could be jointly trained on a mixture of structural data and fact-of-interaction data.
In the context of the Dock2D toy data, the transferability of learned features is due to the fact that the same energy function was used to generate both the Dock2D-IP and Dock2D-IF datasets. While it is unclear to what extent the same approach will work on experimental data, since the connection between configurational energy $E$ and binding free energy $F$ cannot be exactly defined and since what constitutes a positive interaction depends on the experimental assay used~\cite{hunter2016cell}, we expect many of the strategies developed in this work to be applicable.

\section*{Code and data availability}
The code and data related to this work are available at \href{https://github.com/lamoureux-lab/Dock2D}{https://github.com/lamoureux-lab/Dock2D}.

\section*{Acknowledgements}
The authors thank Arth Dharaskar for useful discussions early in the project.
This work has been supported by grant RSF 22-74-10098 to G.D.

\bibliographystyle{unsrt}
\bibliography{citations}

\clearpage
\renewcommand\thefigure{A.\arabic{figure}}
\setcounter{figure}{0}

\newcommand\supw{3.5}
\newcommand\suph{2.1}

\begin{figure}[h]
\includegraphics[width=1.0\linewidth,trim={0 1cm 0 2cm},clip]{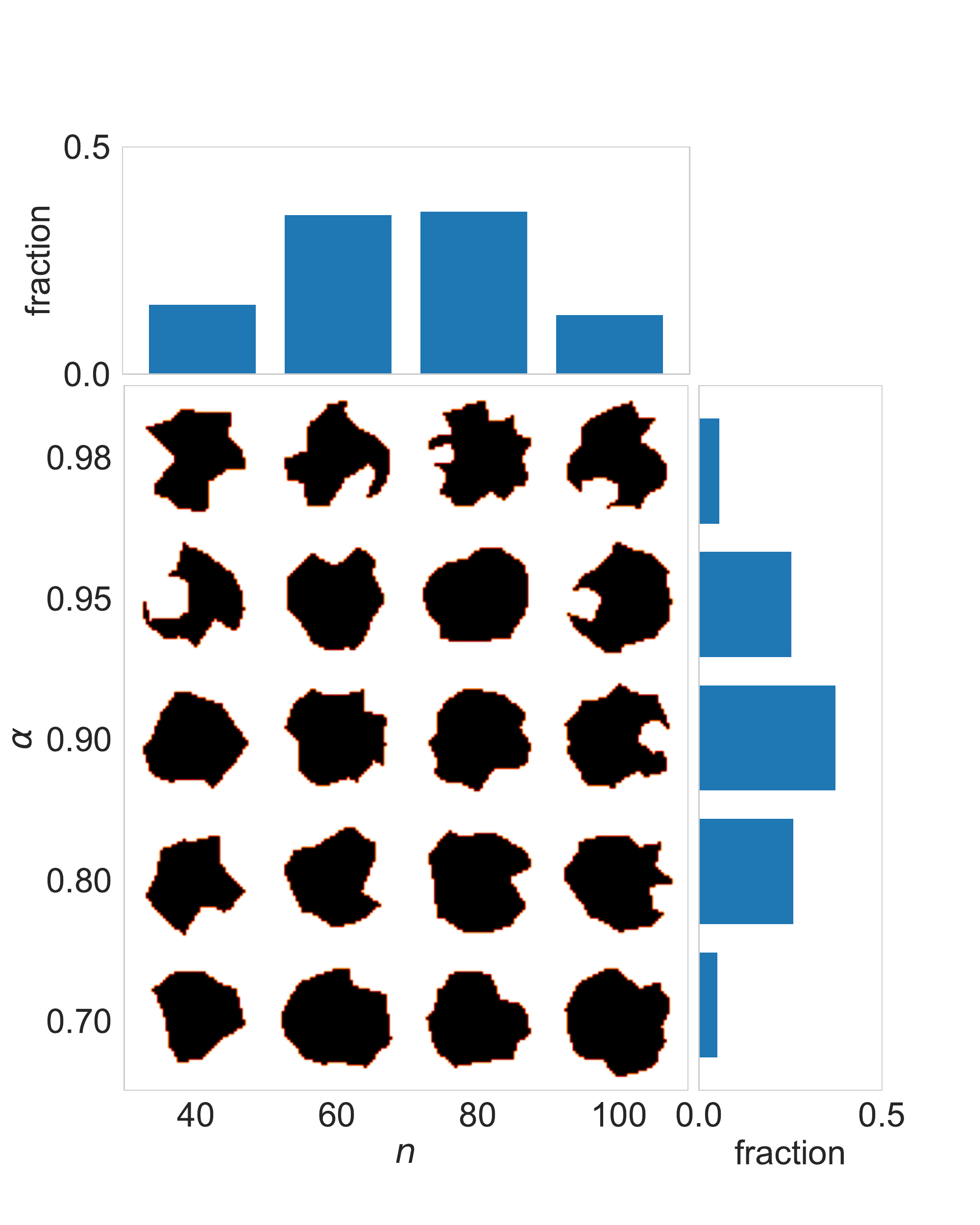}
\caption{Examples of test set shapes generated using predefined values of the $\alpha$ and $n$ parameters. The histograms show how often each value of $\alpha$ and $n$ was picked during shape generation. The $\alpha$ parameter is drawn from probability distribution (0.0625, 0.25, 0.375, 0.25, 0.0625) and the $n$ parameter, from probability distribution (0.125, 0.375, 0.375, 0.125).}
\label{fig:test_shapeDist}
\end{figure}

\begin{figure}[h]
\includegraphics[width=\supw in, height=\suph in,trim={0.5cm 0 0 0},clip]{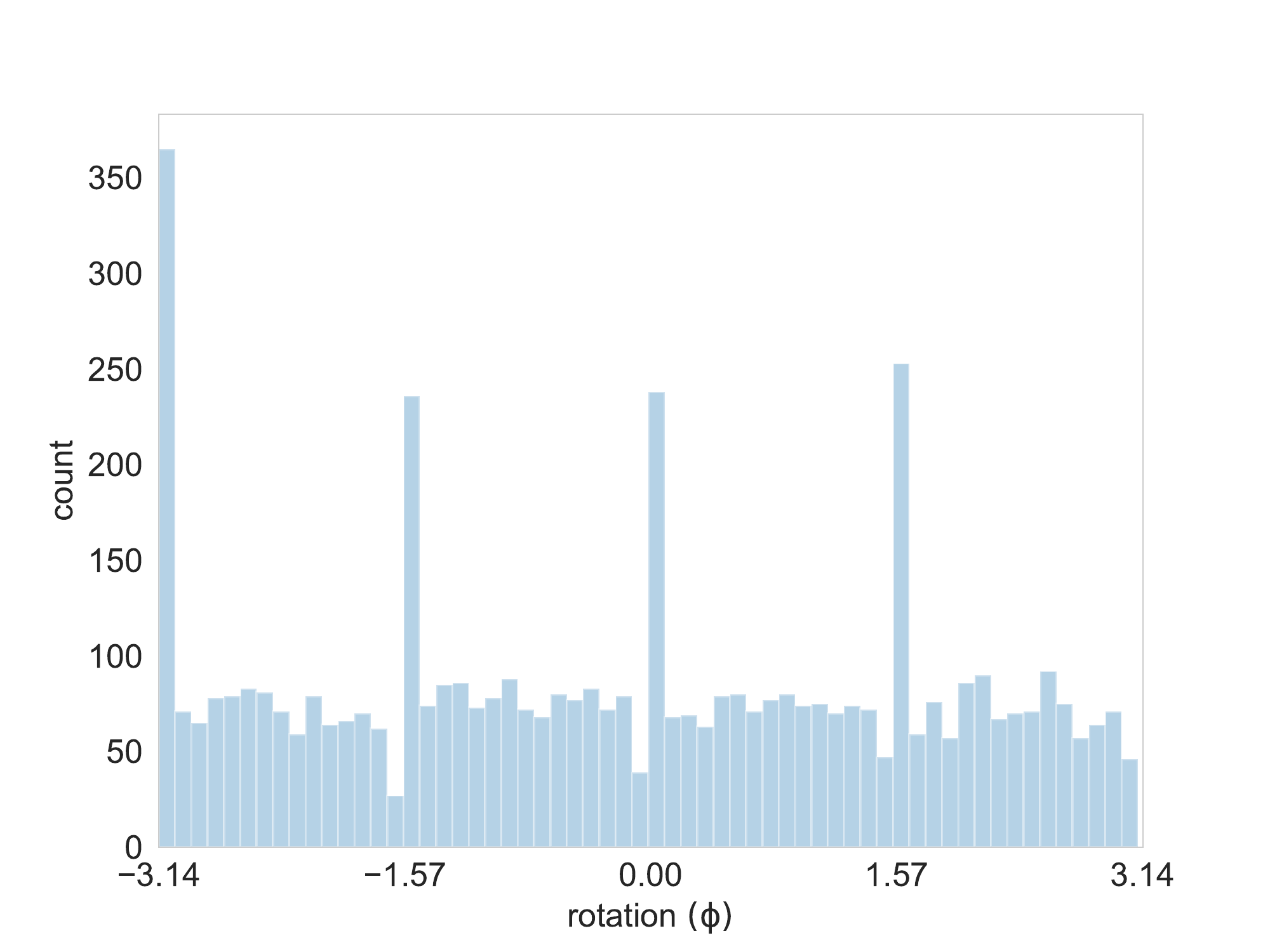}
\includegraphics[width=\supw in, height=\suph in,trim={0.5cm 0 0 0},clip]{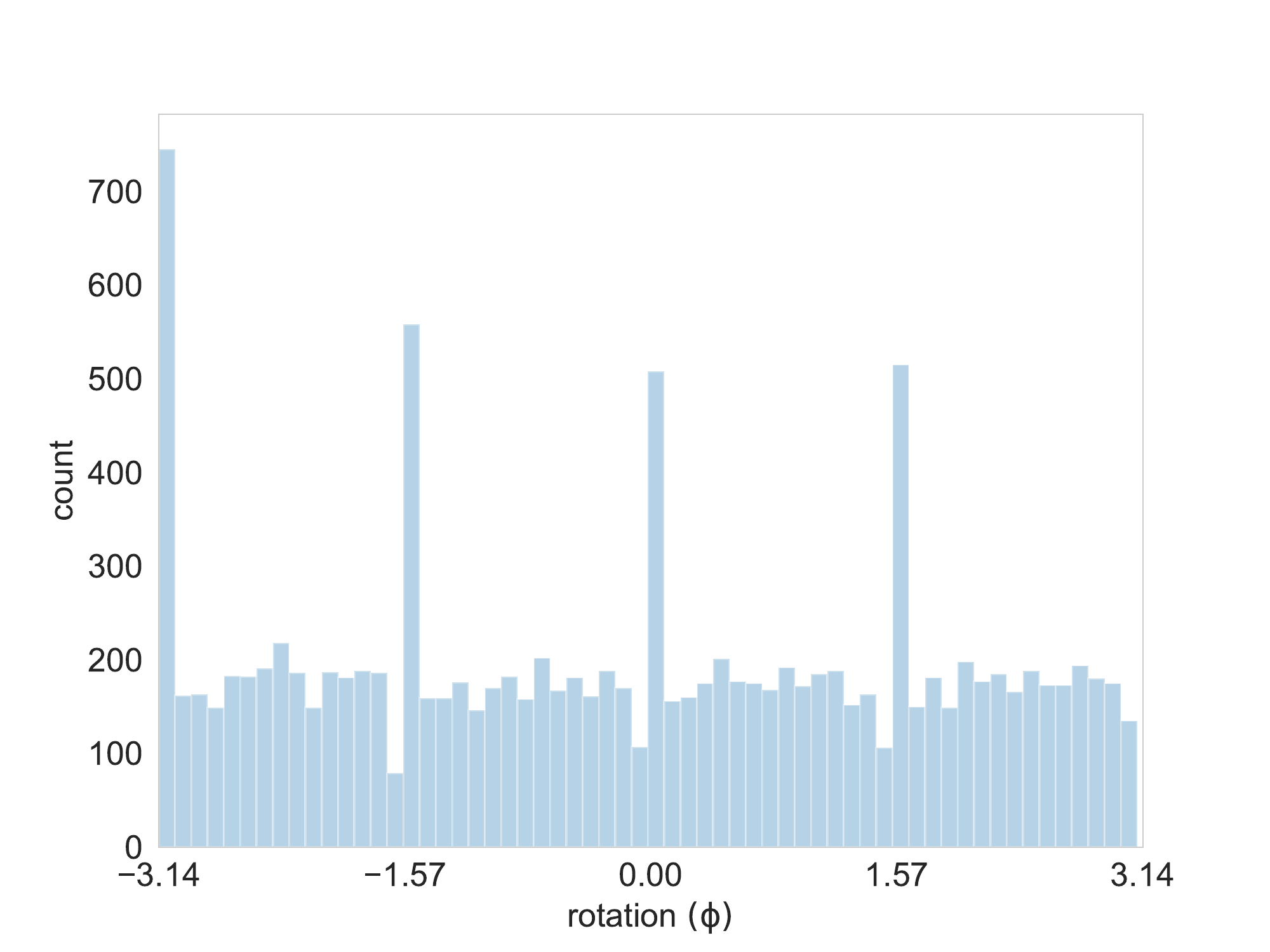}
\caption{Distribution of the ground-truth rotation angle for the Dock2D-IP training/validation set (top panel) and test set (bottom panel).
Rotations by a multiple of $\frac{\pi}{2}$ are more common than expected because the pixelated shapes retain sharper edges when rotated by a multiple of $90^\circ$. Rotation by $-\pi$ ($180^\circ$) is even more common because it is the only possible rotation angle for homodimers.}
\label{fig:sup_rotation_dists}
\end{figure}

\begin{figure}[h]
\includegraphics[width=\supw in, height=\suph in,trim={0 0 0 0},clip]{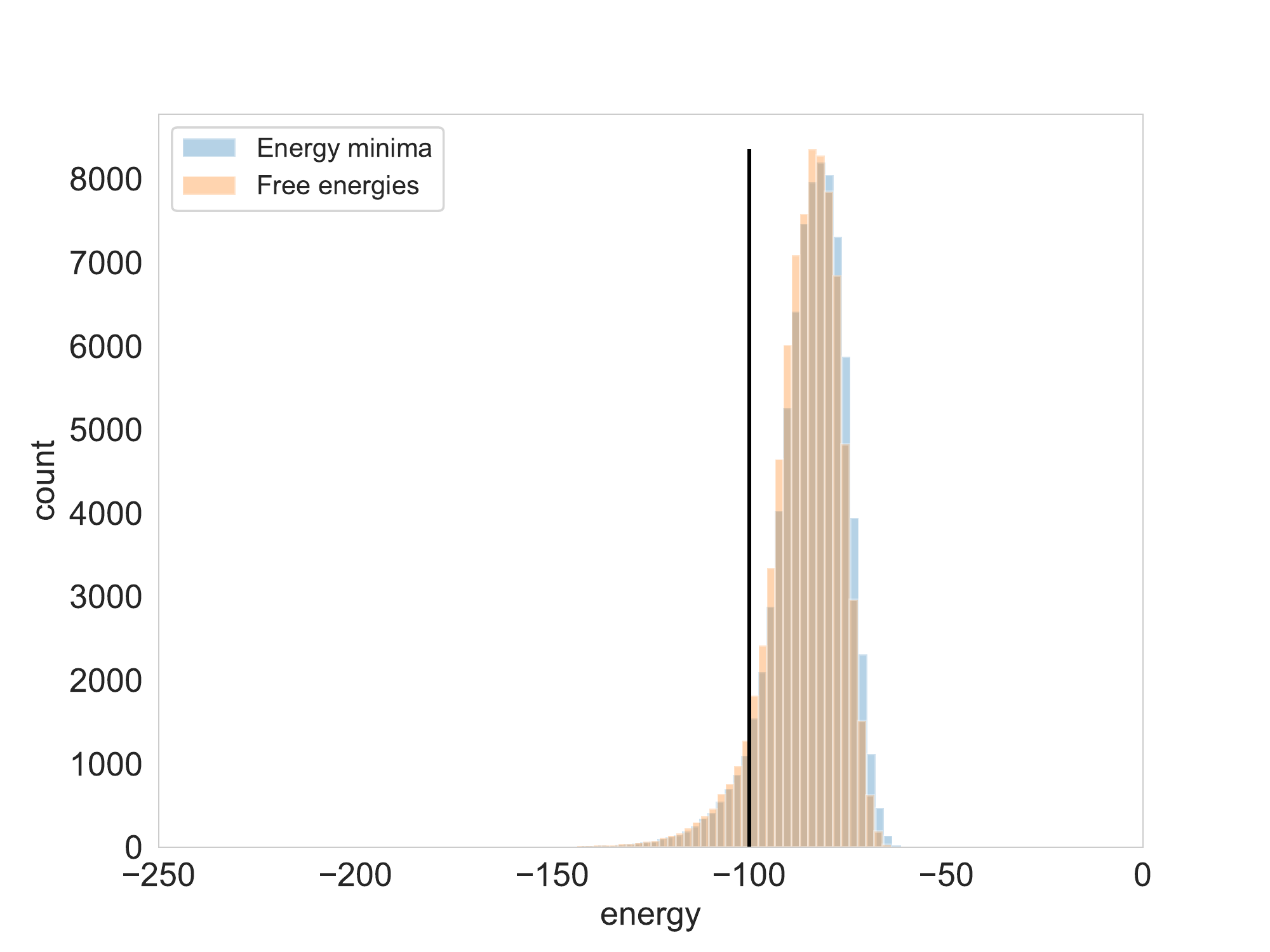}
\includegraphics[width=\supw in, height=\suph in,trim={0 0 0 0},clip]{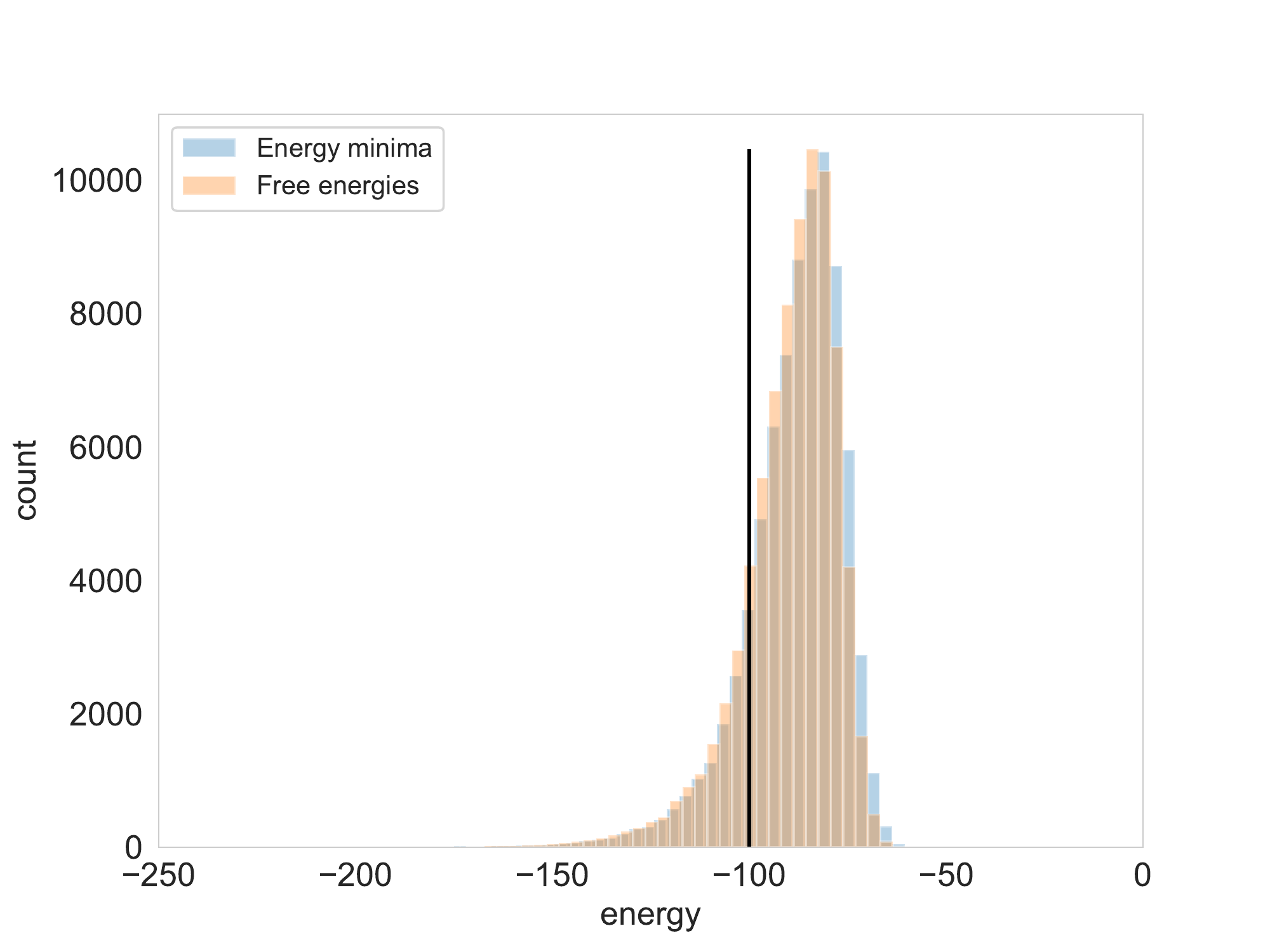}
\caption{Distributions of minimum interaction energy $E_0$ and binding free energy $F$ for pairs of shapes from the training/validation pool (top panel) and the test pool (bottom panel). For the Dock2D-IP set, examples are selected when $E_0 < -100$; For the Dock2D-IF set, they are positive when $F < -100$ and negative when $F > -100$.}
\label{fig:sup_energy_dists}
\end{figure}

\begin{figure}[h]
\includegraphics[
width=1.0\linewidth,trim={7cm 5cm 5cm 2cm},clip]{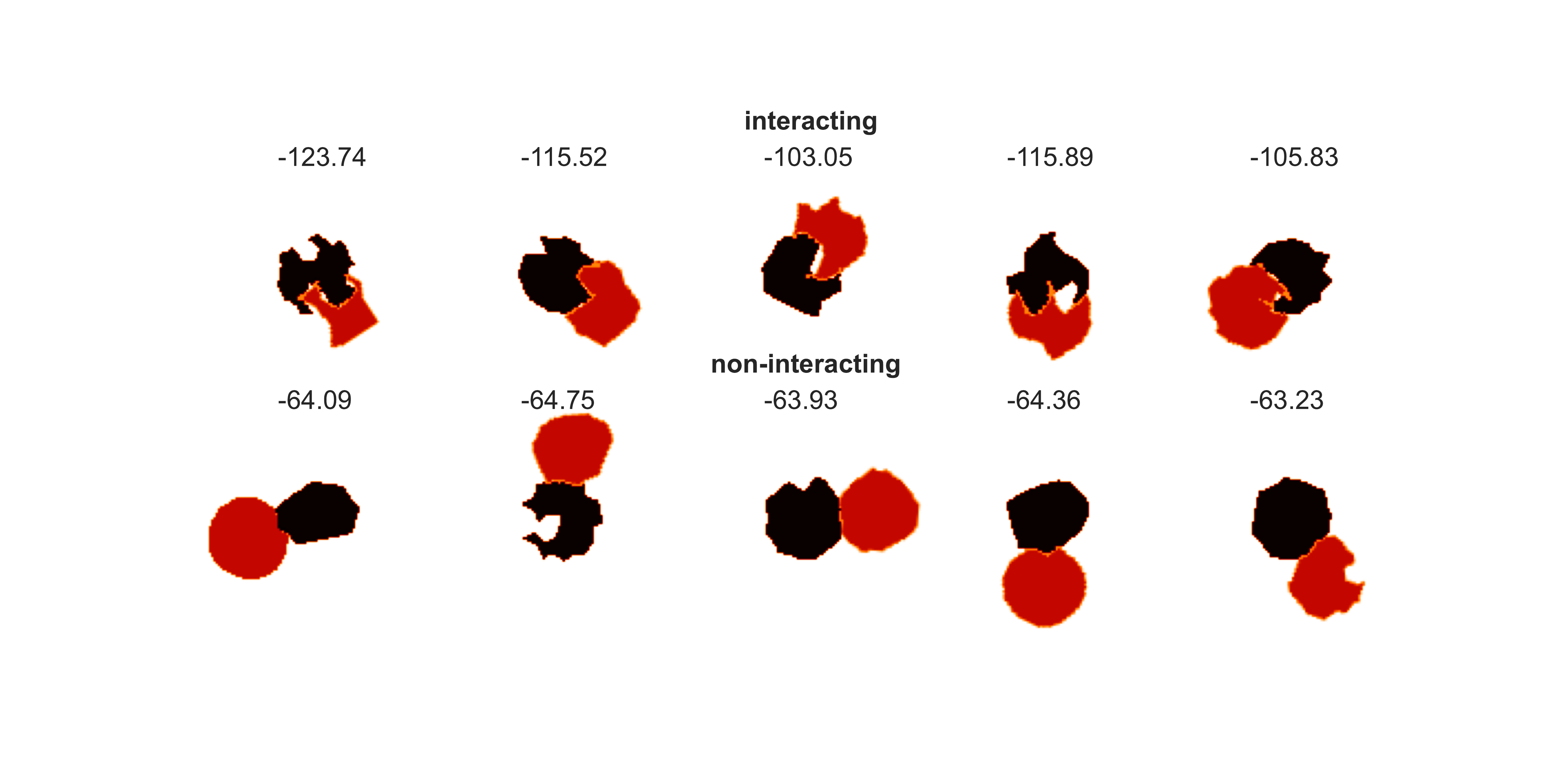}
\caption{Examples of lowest-energy conformations of shapes from the test set forming a positive interaction (top row) and a negative interaction (bottom row). The number above each conformation is the binding free energy $F$.}
\label{fig:interacting_noninteracting_examples_test}
\end{figure}

\begin{figure}[h]
\includegraphics[width=\linewidth, trim={0 0 0 0},clip]{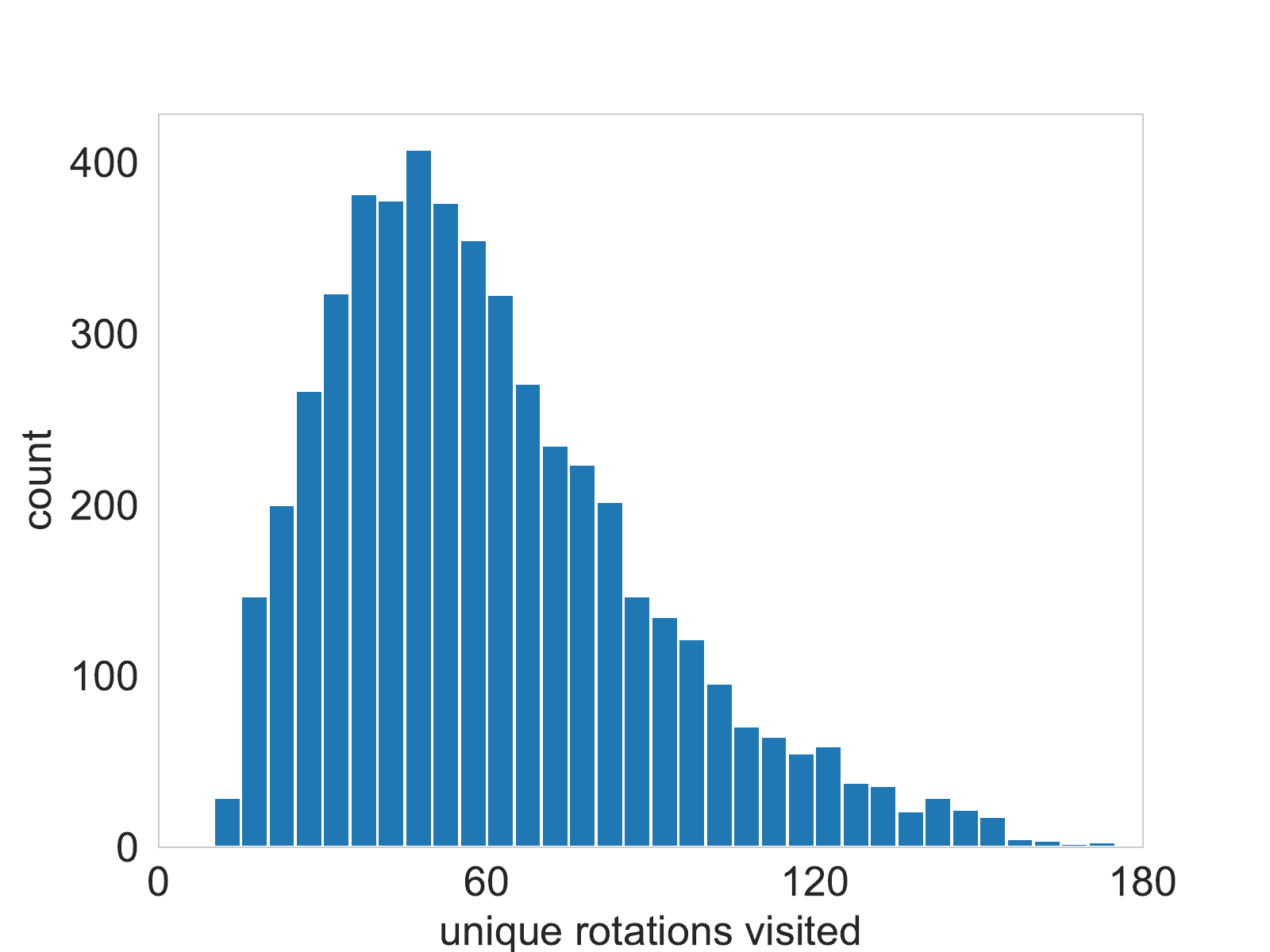}
\caption{Distribution of the number of unique rotations visited (out of a maximum of 180) at the end 1000 epochs of training of the ``sampled IF'' model, for 5050 training examples (100 interaction pairs).}
\label{fig:rotation_surface_saturation}
\end{figure}

\begin{figure}[h]
\includegraphics[width=\supw in, height=\suph in,trim={1.8cm 0 0 0},clip]{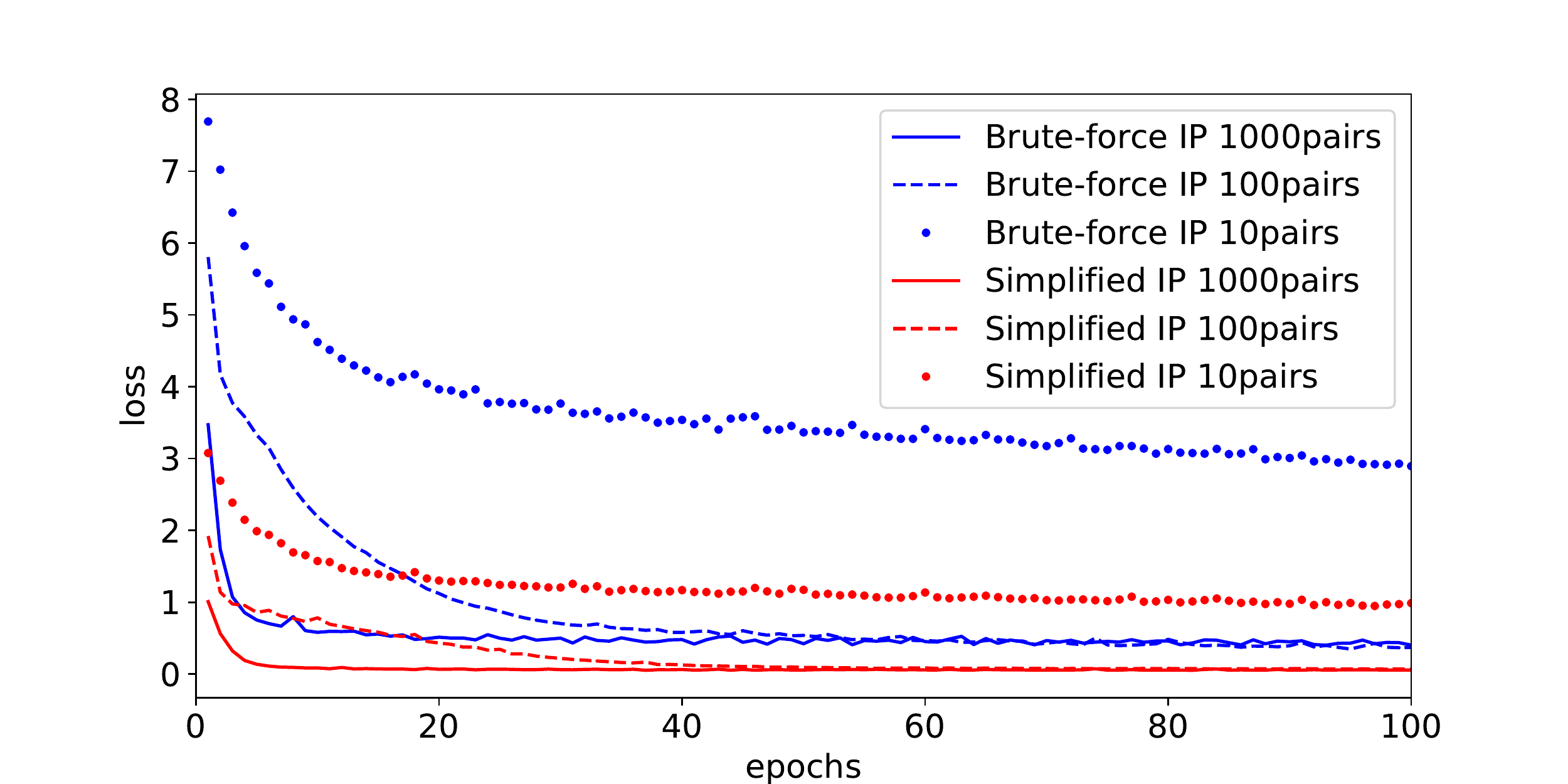}
\caption{Training curves of the models presented in Table~\ref{tbl:IPResults}. For the brute-force models (in blue), cross-entropy loss $\cal{L}_\mathrm{IP}$ is calculated for distribution (\ref{eq:boltzmann_dist}); for the simplified models (in red), it is calculated for distribution (\ref{eq:boltzmann_dist_simplified}). All models were trained for 100 epochs.}
\label{fig:sup_IP_training_curves}
\end{figure}

\begin{figure*}[h]
\begin{center}
\includegraphics[width=\linewidth,trim={0 0 0 0},clip]{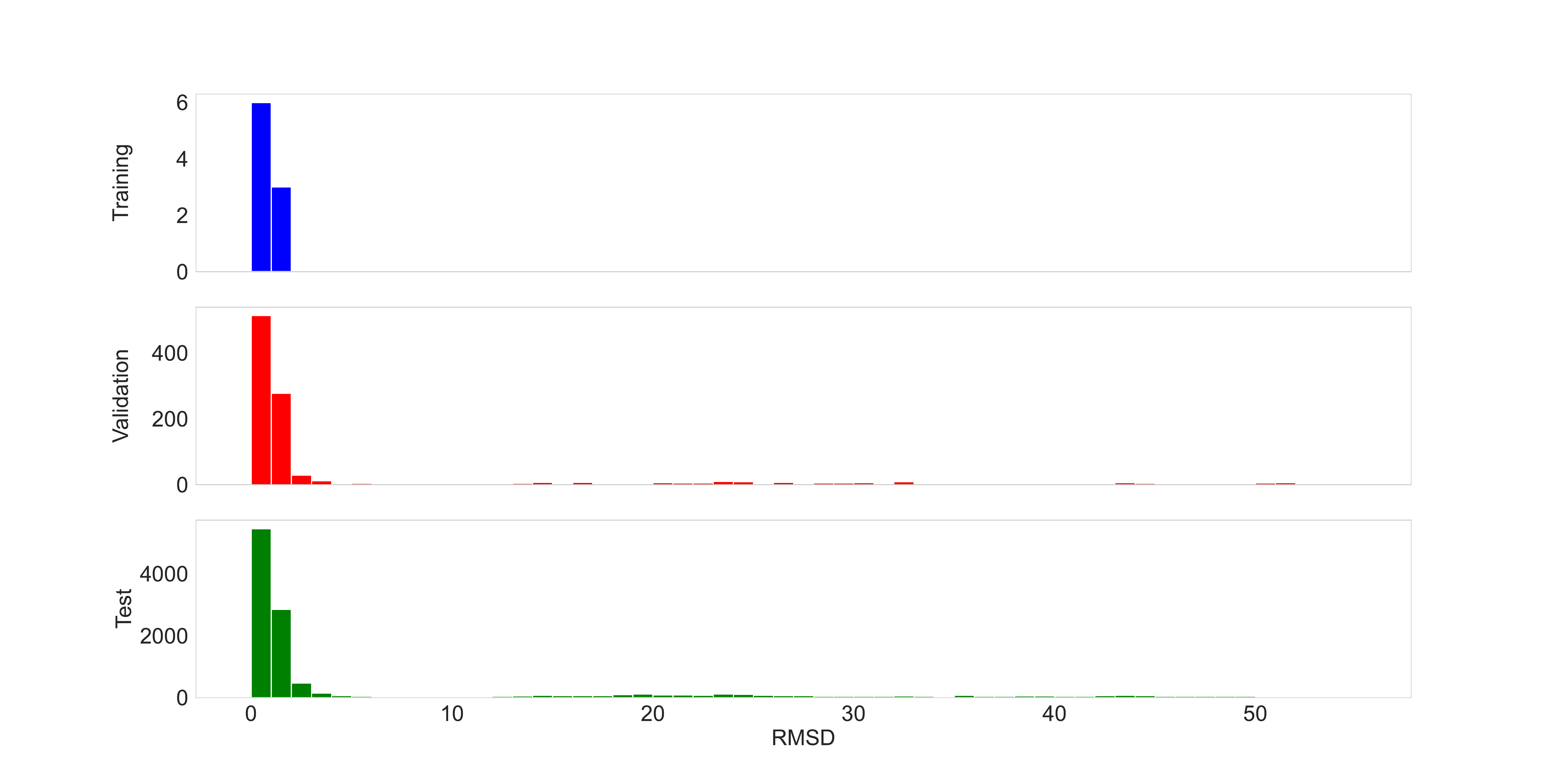}
\caption{Distribution of ligand RMSD for the experiment from Table~\ref{tbl:IPResults} yielding the poorest scores (training on the ``simplified'' IP task using only 10 examples). The average ligand RMSD is 0.37 for the training set (10 examples), 5.97 for the validation set (1017 examples), and 7.64 for the test set (11850 examples).}
\label{fig:RMSD_histogram_BFIP_10ex}
\end{center}
\end{figure*}

\begin{figure}[h]
\includegraphics[width=\supw in, height=\suph in,trim={1.5cm 0 0 0},clip]{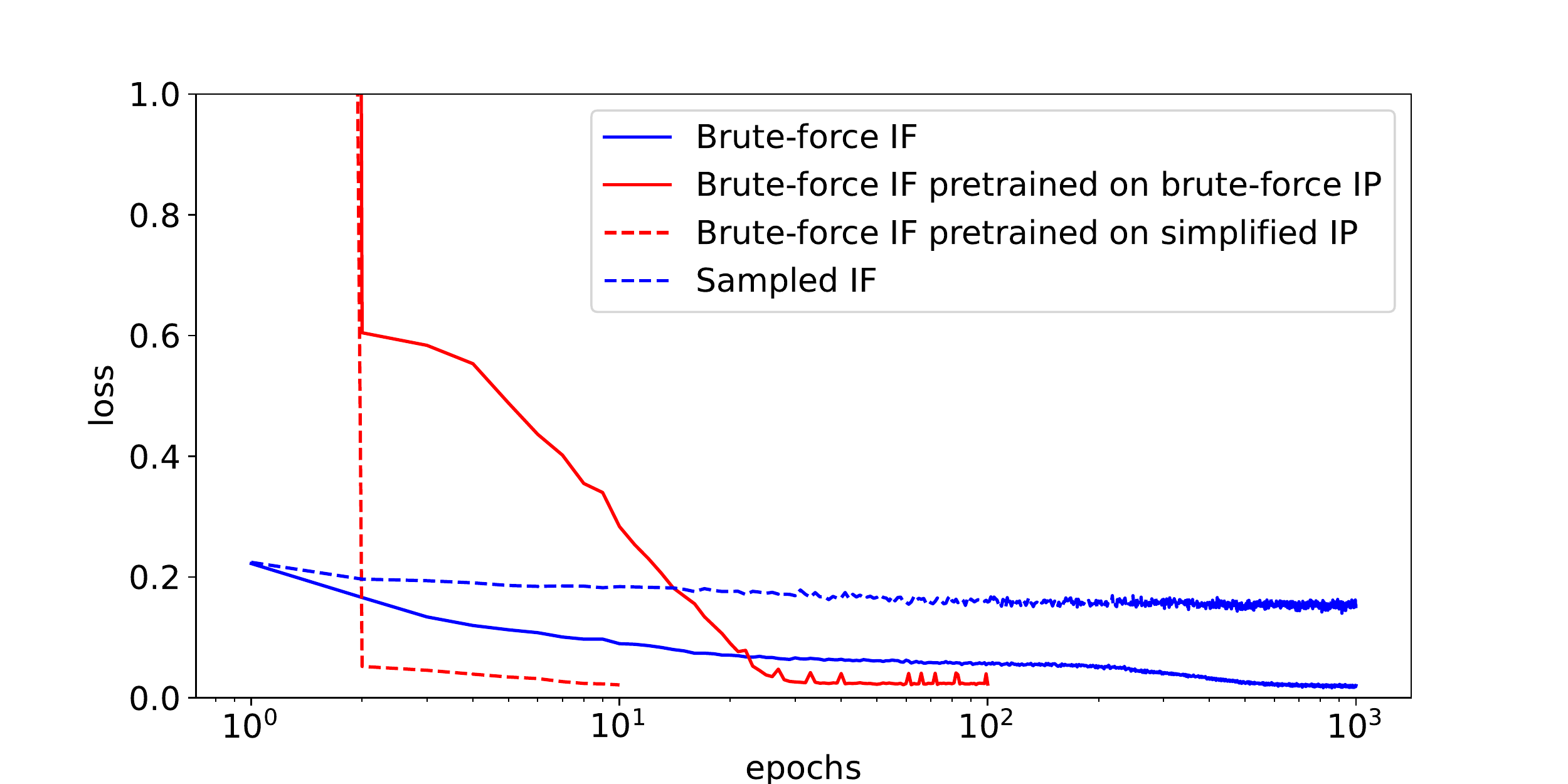}
\caption{Training curves of the models presented in Table~\ref{tbl:IFResults}. The brute-force models pre-trained by transferring the energy function learned from the IP task (red curves) did not require as extensive training.}
\label{fig:sup_IF_training_curves}
\end{figure}

\end{document}